\documentclass[11pt]{article}
\usepackage{epsf}
\usepackage{color}
\usepackage{amsmath,amsfonts,amsbsy,amssymb}
\usepackage{indentfirst}
\usepackage{mathtools}

\newcommand{\fsl}[1]{\ensuremath{\mathrlap{\not{\phantom{#1}}}#1}}

\newcommand{\nn}{\nonumber}

\def\be{\begin{equation}}
\def\ee{\end{equation}}
\def\bse{\begin{subequations}}
\def\ese{\end{subequations}}
\def\bal{\begin{align}}
\def\ealn{\end{align}}
\def\tr{\text{tr}}

\begin{document}

\begin{titlepage}

\def\slash#1{{\rlap{$#1$} \thinspace/}}

\begin{flushright} 

\end{flushright} 

\vspace{0.1cm}

\begin{Large}
\begin{center}

{\bf     Higher Dimensional Quantum Hall Effect \\as\\  A-Class  Topological Insulator
}
\end{center}
\end{Large}

\vspace{0.4cm}

\begin{center}
{\bf Kazuki Hasebe}   \\ 
\vspace{0.3cm} 
\it{Department of Physics, Stanford University, Stanford, California 94305, USA \footnote{On leave from Kagawa National College of Technology, Takuma-cho, Mitoyo, Kagawa 769-1192, Japan. \\
 After 31 March 2014,  email to \sf{hasebe@dg.kagawa-nct.ac.jp}}} \\ 

\vspace{0.4cm} 
{\sf
khasebe@stanford.edu} 

\vspace{0.4cm} 

{\today} 

\end{center}

\vspace{0.2cm}

\begin{abstract}
\noindent

\baselineskip=18pt

We perform a detail study of higher dimensional quantum Hall effects and A-class topological insulators with emphasis on their relations to non-commutative geometry. 
There are two different  formulations of non-commutative geometry for higher dimensional fuzzy spheres; the ordinary commutator formulation and quantum Nambu bracket formulation. Corresponding to these formulations, we introduce 
 two kinds of monopole gauge fields; non-abelian gauge field and antisymmetric tensor  gauge field, which respectively realize the non-commutative geometry of fuzzy sphere in the lowest Landau level.  We establish  connection between the two types of monopole gauge fields through Chern-Simons term, and derive explicit form of tensor monopole gauge fields with higher  string-like singularity.  
 The connection between two types of monopole is applied  
 to  generalize the concept of flux attachment in quantum Hall effect to A-class topological insulator.           
We propose tensor type  Chern-Simons theory as the effective field theory for membranes in  A-class topological insulators.   Membranes turn out to be fractionally charged objects and 
the phase entanglement mediated by tensor gauge field transforms the  membrane statistics to be anyonic.    
The index theorem supports the dimensional hierarchy of A-class topological insulator.  Analogies to D-brane physics of string theory are discussed too.  

\end{abstract}

\end{titlepage}

\newpage 

\tableofcontents

\newpage

\section{Introduction}

 About a decade ago, the time reversal symmetric counterpart of quantum Hall effect,  quantum spin Hall effect, was  theoretically proposed and experimentally discovered  \cite{KaneMele2005,BernevigZhang2005,BernevigHZ2006,Koenigetal2007}. Since then,  topological states of matter have been vigorously investigated [see Refs.\cite{QiandZhangPhysToday,Hasan-Kane-10,Qi-Zhang-11} as reviews].   Now, we understand there exist a variety of topological cousins of quantum Hall effect, such as topological insulators with time reversal symmetry and topological superconductors with particle hole symmetry.     
Based on a generalized Altland and Zirnbauer  random matrix, a systematic classification of the band topological insulators was exhausted in the topological periodic table of ten-fold way  \cite{SchnyderRFL2008,Kitaev2008,RyuSFL2009,QiHZ2008}, where we readily find topological insulators in any dimension with or without three discrete symmetries, time reversal, particle-hole, and chiral. For instance, the quantum Hall effect is assigned to the lowest dimensional (2D)  entity of the A-class topological insulators  that do not respect any of the three discrete symmetries and live in arbitrary even dimensional space. 
 The A-class topological insulators are regarded as a higher dimensional counterpart of the quantum Hall effect. 

Recently, several theoretical realizations of fractional version of topological insulators have been proposed \cite{Neupert-S-R-Ch-M-2011,Santos-N-R-Ch-M-2011}, and two groups independently applied  the non-commutative geometry techniques to fractional 
topological insulators \cite{EstienneRB2012,NeupertSRChMRB2012} generalizing the techniques used in 2D quantum Hall effect \cite{Girvin1984,Girvin-M-P-86,Ezawaetal2003,Haldane2011}. In the works,  they proposed quantum Nambu geometry \cite{Nambu1973,CurtrightZachos2003} as  underling mathematics of topological insulators.  
 In particular, close relations between  quantum Nambu bracket in even dimensions and A-class  topological insulator were pointed out in Ref.\cite{EstienneRB2012} where 
 monopole in the momentum space generates the non-commutativity of density operators. 
Since A-class topological insulators are a natural higher dimensional counterpart of quantum Hall effect, A-class topological insulators give a good starting point to see how non-commutative geometry works in topological insulators before discussing  more ``complicated'' topological insulators, such as AII class\footnote{Recently, AII topological insulators with Landau level were constructed in Refs.\cite{LiWu2013, Li-I-Y-W-2012, LiZhangWu2013}.}.     
 Before the discovery of topological insulators, 4D generalization of quantum Hall effect was theoretically proposed in the $SU(2)$ monopole background by Zhang and Hu \cite{ZhangHu2001} as a generalization of the Haldane's quantum Hall effect on two-sphere \cite{haldane1983}.  In general, 
higher dimensional quantum Hall effects are realized in (color) monopole background compatible with the holonomy group of the basemanifold on which the system is defined \cite{KarabaliNair2002,Bernevig2003,HasebeKimura2003}. 
Since there exists magnetic field of monopole,  higher dimensional quantum Hall effects  necessarily break time-reversal symmetry as A-class topological insulators are ought to do.    
The higher dimensional quantum Hall effect can be considered as a realization of A-class topological insulator with Landau levels\footnote{In this sense, the 4D quantum Hall effect was the firstly ``discovered'' higher dimensional  topological insulator.}.  
From this perspective,  we revisit the  higher dimensional quantum Hall effect that is realized on arbitrary even-dimensional sphere \cite{HasebeKimura2003,Hasebe2010}. 
In the set-up of  quantum Hall effect on $S^{2k}$, the $SO(2k)$ non-abelian monopole is adopted, and the system realizes interesting mathematical structures. 
For instance, the non-abelian monopole mathematically corresponds to the sphere-bundle over sphere \cite{Steenrodbook} where the $S^{2k-1}$-bundle over the base manifold $S^{2k}$ gives the $SO(2k)$ structure group.  
In non-commutative geometry point of view, the system can be regarded as a physical set-up of higher dimensional fuzzy sphere in the lowest Landau level\footnote{Such physical description of fuzzy sphere in monopole background is ``consistent'' with the dielectric effect of D-brane \cite{Myers-1999,Kimura2004}.}.   Interestingly, higher dimensional quantum Hall effects are even related to supersymmetry \cite{Hasebe-Kimura2005, Hasebe2005} and  twistor theory \cite{KarabaliNair2002,Mihai-S-T-2004,Hasebe2009}.

Though in the former articles,  the non-abelian monopoles are adopted in the construction of the higher dimensional quantum Hall effect,  there may be another monopole realization. That is  to use antisymmetric $\it{tensor}$ $U(1)$ monopole.  
Tensor $U(1)$ monopole is a monopole \cite{Nepomechie1985, Teitelboim1986} whose gauge group is $U(1)$ but gauge field is $\it{not}$ a vector but an antisymmetric tensor\footnote{Such antisymmetric tensor gauge field is also known as Kalb-Ramond field \cite{KalbRamond1974}.}.  
While the non-abelian monopole corresponds to an extension of the Dirac monopole by increasing the $\it{internal}$ gauge degrees of freedom, the tensor monopole manifests another extension of the Dirac monopole by increasing the $\it{external}$ indices. 
Therefore, there may be two reasonable generalizations of quantum Hall effect, one is based on the non-abelian monopole and the other is based on the tensor monopole.  One may be  immediately inclined to ask the following questions. What does quantum Hall effect in  tensor monopole background look like and what kind of non-commutative geometry will emerge in the lowest Landau level? If higher dimensional quantum Hall effect has two reasonable generalizations, is there any connection between them?   
For such questions, the precedent researches of  non-commutative geometry give a suggestive hint;     
There are two (superficially) different formulations for higher dimensional fuzzy sphere \cite{Jabbari2004, JabbariTorabian2005,DeBellisSS2010}, one of which is the ordinary commutator formulation and the other is the quantum Nambu bracket formulation.  
Inspired by the observation, we establish connection between the non-abelian and tensor monopole and answer to the questions in this work.  

Topological field theory description of the quantum Hall effect \cite{Zhang-H-K-1988,Zhang-1992} has brought  great progress in understanding non-perturbative aspects of quantum Hall effect. The Chern-Simons effective field theory naturally describes  the flux attachment that electron and Chern-Simons fluxes are combined to yield  a ``new particle'' called composite boson \cite{Girvin-M-1987,Read-1989}, and the fractional quantum Hall effect is regarded as a superfluid state of the composite bosons \cite{Zhang-1992}.  
The fundamental object of the A-class topological insulator turns out to be membrane-like objects.  Based on the connection between the non-abelian and tensor monopoles,  
we propose a tensor type Chern-Simons field theory as an effective field theory of the A-class topological insulator. 
Interestingly, 
while we start from the non-abelian quantum mechanics in $(2k+1)$D space-time, the tensor Chern-Simons field theory is defined in $(4k-1)$D space-time. 
Membranes have a fractional charge and obey anyonic statistics.  The ground state of  A-class topological insulators is regarded as a  superfluid state of composite membrane at magic values of the filling factor.   
We discuss dimensional condensation of membranes with emphasis on its relation to brane-democracy of string theory.

\begin{figure}[tbph]\center
\hspace{5cm}\includegraphics*[width=150mm]{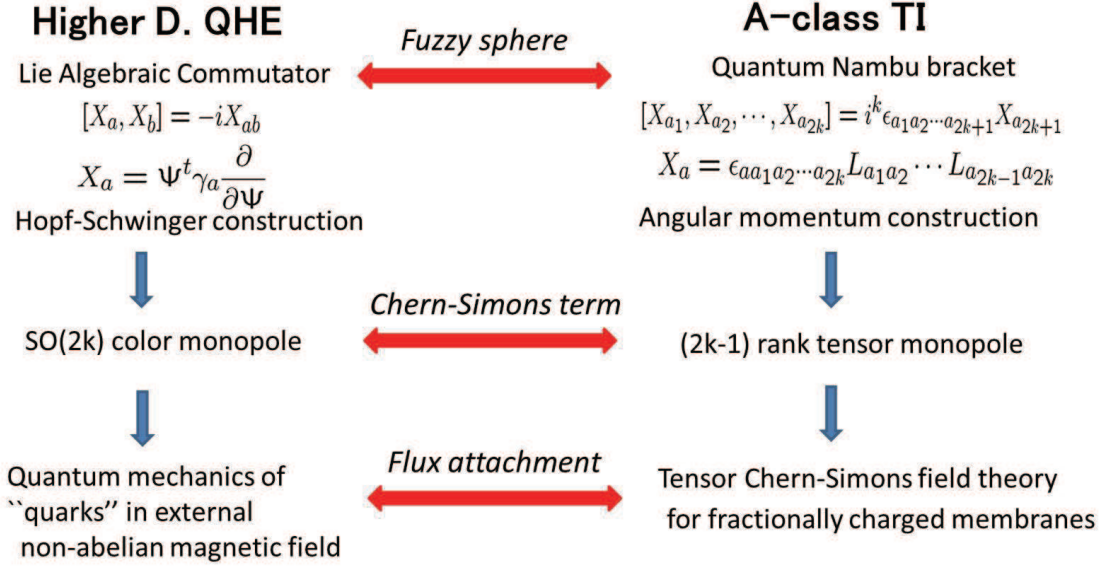}
\caption{Correspondence between mathematics and physics of  higher dimensional quantum Hall effects and A-class topological insulators. }
\label{EntirePicture}
\end{figure}

The main goal of this paper is to integrate so far loosely connected subjects, such as Nambu-bracket, tensor topological field theory and physics of quantum Hall effect, to have an entire picture of  A-class topological insulator [Fig.\ref{EntirePicture}].
 Though we share several terminologies with string theory such as $p$-branes and $C$ field, the present analysis is not directly related to the string theory: We do not use either strings or D-branes. About a realization of topological insulators in string theory,  one may   
 consult Refs.\cite{Ryu-Takayanagi-2010,Furusaki-etal-2012}. For $C$ field realization of non-commutative geometry on M-brane, see Refs.\cite{BergshoeffBSS2000,KawamotoSasakura2000,ChuAhmbi2012}.

The paper is organized as follows. 
In Sec.\ref{sec:fuzzysphereDiracmono}, we briefly review the basic mathematics of the fuzzy sphere and its physical realization in the lowest Landau level.   Sec.\ref{sec:higherdimNCG} describes the two mathematical formulations for higher dimensional fuzzy spheres.   
We introduce non-abelian monopole quantum Hall effect with or without spin degrees of freedom in Sec.\ref{sec:nonabelimonohigherDQHE}.  Sec.\ref{sec:tensormono} discusses the connection  between the tensor and non-abelian monopoles, 
and  gives a tensor monopole realization of the quantum Nambu geometry.  
In Sec.\ref{sec:cstheorymemb},  
the Chern-Simons tensor field theory is proposed as an effective field theory of A-class topological insulator, where we clarify  the fractional charge and anyonic statistics of membranes. 
We also discuss the hierarchical property of membranes and A-class topological insulator.   
Sec.\ref{sec:summary} is devoted to summary and discussions.

\section{Fuzzy Sphere and Dirac Monopole}\label{sec:fuzzysphereDiracmono}

Here, we briefly review how the fuzzy geometry emerges in the context of the lowest Landau level physics by using the fuzzy two-sphere and  Dirac monopole system.  The observation will be a template for higher dimensional fuzzy sphere in the subsequent sections. 
 
The fuzzy two-sphere 
\cite{berezin1975,Hoppe1982,madore1992} is a fuzzy manifold whose coordinates $X_i$ ($i=1,2,3$) satisfy  the $SU(2)$ algebra: 
\be
[X_i,X_j]=i\alpha \epsilon_{ijk}X_k, 
\label{fuzzytwospheresu2}
\ee
and 
\be
X_iX_i=\biggl(\frac{\alpha}{2}\biggr)^2 I(I+2)=r^2(1+\frac{2}{I}). 
\label{fuzzytwosphereradius2}
\ee
Here, $\alpha$ is the unit of non-commutative length and $I$ (integer) specifies the radius of the fuzzy two-sphere $r$ as   
\be
r=\frac{\alpha}{2} I. 
\label{defofalpha} 
\ee
The fuzzy sphere is realized as the lowest Landau level physics.   
We will show how fuzzy geometry emerges on a two-sphere in Dirac  monopole background both from the Lagrange and Hamilton formalisms.  

\subsection{Hopf map and Lagrange formalism}\label{subsec:hopflag}

The Lagrangian for the electron on a two-sphere in monopole background is given by 
\be
L=\frac{M}{2}\dot{x}_i\dot{x}_i-\dot{x}_iA_i, 
\label{lagrangianundermag}
\ee
where $x_i$ $(i=1,2,3)$ are subject to a constraint   
\be
x_ix_i=r^2, 
\ee
and $A_i$ denote the Dirac monopole gauge field  
\be
A_i=-\frac{I}{2r(r+x_3)}\epsilon_{ij3}x_j, 
\label{u1monopolegauge}
\ee
with Dirac monopole charge $I/2$ ($I$ integer) \cite{Dirac1931}.  Relation to the non-commutative geometry will be transparent by introducing the Hopf spinor. 
 The Hopf spinor is the two-component spinor that induces the (1st) Hopf map $S^3~\overset{S^1}\rightarrow~ S^2$:   
\be
\phi ~\rightarrow ~x_i=\frac{\alpha}{2}\phi^{\dagger}\sigma_i\phi,  
\label{1sthopfexp}
\ee
with 
\be
\phi^{\dagger}\phi=I. 
\label{condhopfspi}
\ee
$x_i$ (\ref{1sthopfexp}) automatically satisfy the condition of two-sphere: 
\be
x_ix_i=\biggl(\frac{\alpha}{2}\biggr)^2(\phi^{\dagger}\phi)^2=r^2. 
\ee
The Hopf spinor $\phi$ takes the form 
\be
\phi={\sqrt{\frac{I}{2r(r+x_3)}}}
\begin{pmatrix}
r+x_3 \\
x_1+ix_2 
\end{pmatrix} e^{i\chi}
\ee
with  $e^{i\chi}$ denoting $U(1)$ phase factor, and the monopole gauge field (\ref{u1monopolegauge}) can be derived as   
\be
A=A_idx_i=-i\phi^{\dagger}d\phi.  
\ee
In the lowest Landau level, the kinetic energy is quenched and the Lagrangian (\ref{lagrangianundermag}) is reduced to the following form:   
\be
L_{\text{LLL}}=-A_i \dot{x}_i=i\phi^{\dagger}\frac{d}{dt}\phi.  
\label{loewetLLLag}
\ee
We regard the Hopf spinor as the fundamental variable and derive the canonical momentum of $\phi$ as $i\phi^*$ from (\ref{loewetLLLag}) to apply the quantization condition:   
\be
[\phi_{\alpha},{\phi_{\beta}}^*]=\delta_{\alpha\beta}. 
\ee
After the quantization, the Hopf spinor becomes to the Schwinger operator of harmonic oscillator expressed as\footnote{We can derive the same result in the Hamilton formalism. The lowest Landau level eigenstates are given by the holomorphic function of $\phi$, and its complex conjugate is  effectively represented by the derivative of $\phi$.}     
\be
{\phi_{\alpha}},~{\phi_{\beta}}^*~\rightarrow~\frac{\partial}{\partial\phi_{\alpha}},~\phi_{\beta}, 
\ee
and the coordinates on a two-sphere (\ref{1sthopfexp}) turn out to be  the following operators  
\be
X_i=\frac{\alpha}{2}\phi^t \sigma_i\frac{\partial}{\partial \phi},  
\label{hopfshwingercons}
\ee
which satisfy the fuzzy two-sphere algebra (\ref{fuzzytwospheresu2}), 
and the condition (\ref{condhopfspi}) is rewritten as  
\be
\phi^t\frac{\partial}{\partial\phi}=I.  
\label{constraintI}
\ee
One can readily show that 
Eq.(\ref{hopfshwingercons}) with(\ref{constraintI}) indeed satisfies (\ref{fuzzytwosphereradius2}). 
The emergence of fuzzy sphere is based on the Hopf-Schwinger operator and the Pauli matrices in the Lagrange formalism.

\subsection{Hamilton formalism and angular momentum}\label{subsec:hamil-angl}

The 3D Hamiltonian for a particle in gauge field is generally given by 
\be
H=-\frac{1}{2M}{D_i}^2=-\frac{1}{2M}\frac{\partial^2 }{\partial r^2}-\frac{1}{Mr}\frac{\partial}{\partial r}+\frac{1}{2M r^2}{\Lambda_i}^2, 
\ee
where $D_i$ represent the covariant derivative: 
\be
D_i=\partial_i+iA_i, 
\ee
and $\Lambda_{i}$ denote the covariant angular momentum:  
\be
\Lambda_{i}=-i\epsilon_{ijk}x_jD_k. 
\ee
The Hamiltonian for a particle on two-sphere ($r$ const.) is given by  
\be
H=\frac{1}{2Mr^2}{\Lambda_{i}}^2. 
\ee
With the $U(1)$ monopole at the center of the sphere, the total angular momentum $L_{i}$ is given by the sum of the covariant angular momentum 
and the angular momentum of the monopole gauge field: 
\be
L_{i}=\Lambda_i+r^2F_i=\Lambda_i+\frac{1}{\alpha}x_i, 
\label{rellaandfa}
\ee
where 
\be
F_{i}=\epsilon_{ijk}\partial_j A_k=\frac{I}{2r^3}x_i.   
\ee
Since $L_i$ are the conserved angular momentum, they satisfy the $SU(2)$ algebra 
\be
[L_i, L_j]=i\epsilon_{ijk}L_k. 
\ee
In the lowest Landau level, the kinetic term is quenched $\Lambda_{i}=0$, and then $x_i~(\propto ~F_i)$ can be identified with $L_i$:  
\be
X_i=\alpha L_i.  
\label{xiliequiv}
\ee
It is obvious that $X_i$ satisfy the fuzzy two-sphere algebra (\ref{fuzzytwospheresu2}). 
With use of $L_{ij}=\epsilon_{ijk}L_k$, (\ref{xiliequiv}) is written as  
\be
X_i= \frac{\alpha}{2}\epsilon_{ijk}L_{jk}. 
\label{angularfromxa}
\ee
Notice the construction of fuzzy sphere coordinates in the Hamilton formalism is  based on the angular momentum.  
 
Consequently, there are two ways to see the emergence of fuzzy sphere, one of which is the Hopf-Schwinger construction (\ref{hopfshwingercons}) in the Lagrange formalism, and the other is the angular momentum construction (\ref{angularfromxa}) in the Hamilton formalism. 

\section{Non-commutative Geometry in Higher Dimensions}\label{sec:higherdimNCG}

\subsection{Fuzzy sphere algebra}

As discussed above, the coordinates of fuzzy two-sphere are given by the $SO(3)$ vector operators that satisfy 
\be
[X_i,X_j]=i\alpha \epsilon_{ijk}X_k, \nn\\ 
\ee
and its minimal representation is the $2\times 2$ Pauli matrices. 
Since Pauli matrices are equal to the $SO(3)$ gamma matrices, it may be natural to adopt  the $SO(2k+1)$ gamma matrices as the coordinates of  $S^{2k}_F$ with minimum radius. For $S_F^{2k}$ with larger radius, the $SO(2k+1)$ gamma matrices $G_a$ $(a=1,2,\cdots, 2k+1)$ of fully symmetric representation\footnote{For several properties of gamma matrix in fully symmetric representation, see Append.\ref{append:so2k+1gammagenefulsym}.},  $\overbrace{[\frac{I}{2},\frac{I}{2}\cdots,\frac{I}{2}]}^{k}$, is adopted as the fuzzy coordinates\cite{Grosse-K-P-1996,Kabat-Taylor-1998}. Indeed  $X_a\equiv\alpha G_a$ satisfy 
\be
\sum_{a=1}^{2k+1}X_aX_a=\frac{\alpha^2}{4}I(I+2k)=r^2(1+\frac{2k}{I}), 
\label{fuzzyspherecondhigh}
\ee
which represents the condition of constant radius of fuzzy sphere.  In the limit $I\rightarrow \infty$ with fixed $r$, (\ref{fuzzyspherecondhigh}) is reduced to the classical condition of $2k$-sphere,  $\sum_{a=1}^{2k+1}x_ax_a=r^2$. 

One should notice however, there is a big difference between the fuzzy two-sphere and its higher dimensional counterpart \cite{Ramgoolam2001, HoRamgoolam2002, Kimura2002, 
Kimura2003}. Though the $SO(3)$ gamma matrices are equivalent to the $SU(2)$ generators and form a closed algebra by themselves, the $SO(2k+1)$ $(k\ge 2)$ gamma matrices $X_a$ do not satisfy a closed algebra among themselves but their commutators yield ``new'' operators,  the $SO(2k+1)$ generators $X_{ab}$:  
\be
[X_{a},X_{b}]=i\alpha X_{ab}. 
\label{commufuzzycoords}
\ee
The appearance of $X_{ab}$ suggests that the geometry of higher dimensional fuzzy sphere cannot simply be understood only by the original coordinates.      
To construct a closed algebra for higher dimensional fuzzy sphere, we need to incorporate $X_{ab}$ to have an enlarged  algebra 
\begin{align}
&[X_a,X_{bc}]= -i\alpha(\delta_{ab}X_c-\delta_{ac}X_b)  ,\nn\\
&[X_{ab},X_{cd}]= i\alpha(\delta_{ac}X_{bd}-\delta_{ad}X_{bc}+\delta_{bd}X_{ac}-\delta_{bc}X_{ad}), 
\label{enlargealgebra}
\end{align}
in which $X_a$ and $X_{ab}$ amount to the $SO(2k+2)$ algebra.   
Around the north pole, (\ref{commufuzzycoords})  reduces to  
\be
[X_{\mu},X_{\nu}]=i\alpha {\eta_{\mu\nu}}^i X_i, 
\ee
where ${\eta_{\mu\nu}}^i$ denotes the  expansion coefficient (for $k=2$, ${\eta_{\mu\nu}}^i$ is given by the t'Hooft symbol) and $X_i$ stand for the $SO(2k)$ generators related to $X_{\mu\nu}$ by the relation  
\be
X_{\mu\nu}= \sum_{i=1}^{k(2k-1)}{\eta_{\mu\nu}}^i X_i. 
\label{expanstionxmunusxi}
\ee
The extra-degrees of freedom is described by the operators $X_{i}$, and can be interpreted as the fuzzy fibre-bundle over $S^{2k}$. Since the corresponding algebra of $S_F^{2k}$ is the $SO(2k+2)$ algebra, the fuzzy fibre described by the $SO(2k)$ algebra is identified with  $S_F^{2k-2}$.  
Due to the existence of the fuzzy bundle, 
the classical counterpart of $S_F^{2k}$ is not simply given by  $S^{2k}\simeq SO(2k+1)/SO(2k)$ but  $SO(2k)/U(k)$ fibration over $S^{2k}$ \cite{HoRamgoolam2002}: 
\be
S_F^{2k}\simeq SO(2k+1)/U(k)~\sim~S^{2k}\otimes SO(2k)/U(k). 
\label{fuzzysphereproduct}
\ee
Here, $\sim$ denotes the local equivalence.  
The $SO(2k)/U(k)$-fibre is the classical counterpart of the extra fuzzy space $S_F^{2k-2}$.  
 As we shall see later, such extra degrees of freedom correspond to (fuzzy) membrane excitation. 

Though in the  commutator formulation,  the existence of the fuzzy fibre is explicit, the commutator formulation is rather ``awkward'' in the sense the algebra does not close within the original fuzzy coordinates. The Nambu bracket gives a more sophisticated formulation.    
 In the $d$ dimension, quantum Nambu bracket (or Nambu-Heisenberg bracket) \cite{CurtrightZachos2003, Jabbari2004, JabbariTorabian2005, DeBellisSS2010} is defined as  
\be
[X_{a_1},X_{a_2},\cdots,X_{a_n}]\equiv X_{[a_1}X_{a_2}\cdots X_{a_{n}]},  
\label{defofnambubracket}
\ee
where  $a_1,a_2,\cdots, a_n=1,2,\cdots, d$ ($n \le d$)\footnote{For $n > d$, due to the anti-symmetric property, quantum Nambu bracket always vanishes}, and the bracket for the low indices represents the fully anti-symmetric combination about the indices.   We have $n!$ terms on the right-hand side of (\ref{defofnambubracket}). 
For instance, 
\begin{align}
[X_{a_1 } X_{a_2}]&=X_{a_1}X_{a_2}-X_{a_2}X_{a_1},\nn\\
[X_{a_1 } X_{a_2}X_{a_3}]&= X_{a_1}X_{a_2}X_{a_3}-X_{a_1}X_{a_3}X_{a_2}+ X_{a_2}X_{a_3}X_{a_1}-X_{a_2}X_{a_1}X_{a_3}+X_{a_3}X_{a_1}X_{a_2}-X_{a_3}X_{a_2}X_{a_1}\nn
 \end{align}
In the quantum Nambu bracket formulation\footnote{(\ref{fuzzynbracksphere}) essentially comes from  the property of the $SO(2k+1)$ gamma matrices, $\gamma_1\gamma_2\gamma_3 \cdots \gamma_{2k}=i^k \gamma_{2k+1}.$ 
For more detail properties  of quantum Nambu bracket, see Appendix \ref{append:nambu-heisenberg}.}, the non-commutative algebra for $S_F^{2k}$ is given by \cite{Jabbari2004, JabbariTorabian2005, DeBellisSS2010}
\be
[X_{a_1},X_{a_2},X_{a_3},\cdots, X_{a_{2k}}]=i^k C(k,I) \alpha^{2k-1}\epsilon_{a_1 a_2 a_3 \cdots a_{2k+1}}X_{a_{2k+1}}, \label{fuzzynbracksphere}
\ee
where 
\be
C(k,I)=\frac{(2k)!!(I+2k-2)!!}{2^{2k-1}I!!}. 
\label{defofckI}
\ee
Thus, the extra operators $X_{ab}$ do not appear in the quantum Nambu bracket formulation, and the closure of algebra is guaranteed only by the original fuzzy coordinates.  The extra fuzzy-fibre degrees of freedom seem to be completely ``hidden'' in the quantum Nambu bracket. 
Around the north-pole $X_{2k+1}\simeq r $, (\ref{fuzzynbracksphere}) is reduced to the quantum Nambu bracket for the non-commutative plane:  
\be
[X_{\mu_1},X_{\mu_2},X_{\mu_3},\cdots,X_{\mu_{2k}}]=i^k\ell^{2k} \epsilon_{\mu_1 \mu_2 \mu_3 \cdots \mu_{2k}},  
\label{quantumnambudeffuzzysph}
\ee
where 
\be
\ell\equiv \alpha\biggl(\frac{I}{2}C(k,I)\biggr)^{\frac{1}{2k}}={r}
\biggl(\frac{(2k)!!(I+2k-2)!!}{I!! I^{2k-1}}  \biggr)^{\frac{1}{2k}}~~~\overset{I\sim \infty}{\sim}~~~\frac{r}{\sqrt{I}} . 
\ee
For instance, 
\begin{align}
&k=1~~:~~\ell=r\biggl(\frac{2}{I}\biggr)^{\frac{1}{2}}, \nn\\
&k=2~~:~~\ell=r\biggl(\frac{8(I+2)}{I^3}\biggr)^{\frac{1}{4}}, \nn\\
&k=3~~:~~\ell=r\biggl(\frac{48(I+2)(I+4)}{I^5}\biggr)^{\frac{1}{6}}.  
\end{align}

\subsection{Two monopole set-ups for higher dimensional fuzzy sphere}\label{subsec:twomonop}

As discussed in Sec.\ref{sec:fuzzysphereDiracmono}, the fuzzy two-sphere is realized in the Dirac monopole background.  The easiest way to find what kind of monopole corresponds to  non-commutative geometry is to find the right-hand side of the non-commutative algebra. For instance, the fuzzy two-sphere algebra is  given by  
\be
[X_i,X_j]=i\alpha \epsilon_{ijk}X_k,  \\
\ee
and one can read off the  $U(1)$ monopole field strength from its right-hand side:  
\be
F_{ij}\simeq  \frac{1}{r^3}\epsilon_{ijk}x_k. 
\ee
For higher dimensional fuzzy sphere, 
in correspondence to the two non-commutative formulations, we will obtain two different types of monopoles.

\begin{itemize}
\item Non-abelian monopole
\end{itemize}

Around the north pole, the commutation relation between the fuzzy coordinates (\ref{commufuzzycoords}) becomes to  
\be
[X_{\mu},X_{\nu}]=i\alpha X_{\mu\nu}, \nn\\
\ee
where the right-hand side is the $SO(2k)$ generators. 
This suggests the $SO(2k)$ non-abelian monopole field strength:    
\be
F_{\mu\nu}\simeq  \frac{1}{r^2}\Sigma_{\mu\nu}, 
\label{aroudnonfield}
\ee
where $\Sigma_{\mu\nu}$ denotes the $SO(2k)$ matrix generators. 
Thus, 
we can identify one monopole set-up for $S_F^{2k}$ with the  $SO(2k)$ non-abelian monopole.

\begin{itemize}
\item Tensor monopole set-up
\end{itemize}

Meanwhile, the right-hand side of the quantum Nambu bracket formulation  
\be
[X_{a_1},X_{a_2},\cdots, X_{a_{2k}}]=i^{k}C(k,I)\alpha^{2k-1}\epsilon_{a_1 a_2 \cdots a_{2k} a_{2k+1} }X_{a_{2k+1}}, \nn\\
\ee
implies  antisymmetric tensor monopole field strength:  
\be
G_{a_1 a_2 \cdots a_{2k}}\simeq \frac{1}{r^{2k+1}} \epsilon_{a_1 a_2 \cdots a_{2k+1}} x_{a_{2k+1}}. 
\label{fieldstrength2k+1}
\ee
Here two comments are added.   
Firstly, even though there are two different non-commutative formulations, they describe the same non-commutative object, $i.e.$ the fuzzy sphere, and then the two different types of  monopoles are expected to describe same physical system corresponding to fuzzy sphere.   
In other words, the non-abelian and the tensor monopoles are two different physical set-ups for the same  system.  
They are expected to be  ``equal'' in some sense. Their connection will be clarified in Sec.\ref{sec:tensormono}.  
Secondly, 
though the quantum Nambu algebra veils the ``extra'' degrees of freedom of fuzzy-bundle,   $(2k-1)$ rank field (\ref{fieldstrength2k+1}) implies the  
existence of $(2k-2)$-brane whose $(2k-1)$-from current naturally coupled to $(2k-1)$ rank tensor field.  
 This observation will be important in constructing the Chern-Simons tensor field theory in Sec.\ref{sec:cstheorymemb}.

\section{Non-Abelian Monopole and Higher Dimensional Quantum Hall Effect}\label{sec:nonabelimonohigherDQHE}

Here, we give non-abelian monopole realization for higher dimensional quantum Hall effect \cite{HasebeKimura2003, Hasebe2010}.   
The $SO(2k)$ monopole gauge group is adopted so as  to be compatible with the holonomy of the basemanifold $S^{2k}$\footnote{ The present monopole set-up is quite similar to the Kaluza-Klein monopole in the sense that the  geometrical information determines the corresponding monopole gauge group. Kaluza-Klein monopole accompanies with  the spontaneous compactification of the Kaluza-Klein theory \cite{Sorkin1983,GrossPerry1983}, and  the isometry of the compactified space is transfered to the gauge symmetry of the uncompactified space. 
For instance,   $S^{2k-1}$ compactification yields the $SO(2k)$ gauge symmetry of non-Abelian  monopole \cite{Perry1984}.     }.

\subsection{$SO(2k)$ non-Abelian monopole}\label{subsec:so2knonabelian}

First let us introduce the generalized Hopf map:   
\be
x_a =\alpha \Psi^{\dagger}\Gamma_a \Psi, 
\label{genelahopf}
\ee
where $x_a$ $(a=1,2,\cdots, 2k+1)$ are subject to the condition of $S^{2k}$ : 
\be
x_a x_a=r^2, 
\ee
and $\Gamma_a$ $(a=1,2,\cdots, 2k+1)$ denote the $SO(2k+1)$ gamma matrices: 
\be
\Gamma_i=\begin{pmatrix}
0 & i\gamma_i \\
-i\gamma_i & 0 
\end{pmatrix},~~\Gamma_{2k}=
\begin{pmatrix}
0 & \bold{1}_{2^{k-1}} \\
\bold{1}_{2^{k-1}} & 0 
\end{pmatrix},~~\Gamma_{2k+1}=
\begin{pmatrix}
\bold{1}_{2^{k-1}} & 0 \\
0 & -\bold{1}_{2^{k-1}}
\end{pmatrix}, \label{gammamtrso2k+1}
\ee
with $SO(2k-1)$ gamma matrices $\gamma_i$ ($i=1,2,\cdots 2k-1$).  
The $SO(2k)$ generators 
\begin{align}
\Sigma_{\mu\nu}
 \equiv -i\frac{1}{4}[\Gamma_{\mu}, \Gamma_{\nu}]. \label{generatorsso2k+1}
\end{align}
take the form of 
\be
\Sigma_{\mu\nu}=\begin{pmatrix} 
\Sigma^+_{\mu\nu} & 0 \\
0 & \Sigma^-_{\mu\nu}
\end{pmatrix}, 
\label{sigmapmdec}
\ee
where the $SO(2k)$ Weyl generators  are 
\be
\Sigma^{\pm}_{\mu\nu}=\{\Sigma^{\pm}_{ij}, \Sigma^{\pm}_{i,2k}    \}= \{ -i\frac{1}{2}\gamma_i \gamma_j, \pm \frac{1}{2}
\gamma_i\}. ~(i\neq j) 
\label{weylgenes}
\ee
Notice that the $SO(2k)$ Weyl generators (\ref{weylgenes}) consist of the $SO(2k-1)$ generators and the $SO(2k-1)$ gamma matrices.  
The $2^{k}$ component spinor $\Psi$ that satisfies (\ref{genelahopf}) is given by 
\begin{equation}
\Psi=\frac{1}{\sqrt{2r(r+x_{2k+1})}}
\begin{pmatrix}
(r+x_{2k+1})\bold{1}_{2^{k-1}}\\
x_{2k}\bold{1}_{2^{k-1}}-ix_i\gamma_i
\end{pmatrix}\psi, 
\label{geneHopf}
\end{equation}
where $\psi$ is a $2^{k-1}$ component normalized complex spinor $\psi^{\dagger}\psi=I$. 
With use of $\Psi$,  the $SO(2k)$ non-abelian gauge fields \cite{Yang1978, HorvathPalla1978, Tchrakian1980, Grossmanetal1984, Saclioglu1986} can be derived by the formula 
\begin{equation}
A=-i\Psi^{\dagger}d\Psi,  
\end{equation}
where  $A=A_{a}dx_a$ with   
\begin{align}
&A_{\mu}=-\frac{1}{r(r+x_{2k+1})}\Sigma_{\mu\nu}^+x_{\nu},~~(\mu,\nu=1,2,\cdots,2k),\nn\\
&A_{2k+1}=0.\label{monopolegaugefield}
\end{align} 
The field strength $F=dA+iA^2$ or  $F_{ab}=\partial_a A_b-\partial_bA_a+i[A_a,A_b]$ 
($F=\frac{1}{2}F_{ab}dx_a \wedge dx_b$) is evaluated as 
\footnote{ The component fields of 
$A_a$ and $F_{ab}$ are respectively given by 
\begin{align}
A_{a}=\sum_{\mu<\nu}A_{a}^{~~\mu\nu}\Sigma_{\mu\nu}^+, ~~~~~F_{ab}=\sum_{\mu<\nu}F_{ab}^{~~\mu\nu}\Sigma_{\mu\nu}^+,
\end{align}
where 
\begin{equation}
A_{a}^{~~\mu\nu}=-\frac{1}{r(r+x_{2k+1})}(\delta_{a\mu} x_{\nu}-\delta_{a\nu}x_{\mu}).
\end{equation}
and 
\begin{align}
&F_{\rho\sigma}^{~~\mu\nu}
=\frac{1}{r^3(r+x_{2k+1})}(\delta_{\rho\mu} x_{\sigma} x_{\nu}-\delta_{\rho\nu}x_\sigma x_{\mu}+\delta_{b\mu} x_\rho x_{\nu}-\delta_{\sigma\nu}x_\rho x_{\mu})+\frac{1}{r^2}(\delta_{\rho\mu}\delta_{\sigma\nu}-\delta_{\rho\nu}\delta_{\sigma\mu}),\nn\\
&F_{\rho,2k+1}^{~~\mu\nu}
=-\frac{1}{r^3}(\delta_{\rho\mu}x_{\nu}-\delta_{\rho\nu}x_{\mu}).
\end{align}\label{monopolefieldstrength}
} 
\begin{align}
&F_{\mu\nu}=-\frac{1}{r^2}x_{\mu}A_{\nu}+\frac{1}{r^2}x_{\nu}A_{\mu}+\frac{1}{r^2}\Sigma_{\mu\nu}^+,\nn\\
&F_{ \mu, 2k+1}=\frac{1}{r^2}(r+x_{2k+1})A_{\mu}.\label{nonabelstrength}
\end{align}
Around the north pole, $x_{2k+1}/r\simeq 1$ $x_{\mu}/r\simeq 0$, the field strength (\ref{nonabelstrength}) is reduced to (\ref{aroudnonfield}). 
It is obvious that under the  $SO(2k)$ gauge transformation 
\begin{equation}
\Psi~\rightarrow~
\begin{pmatrix} 
g & 0 \\
0 & g
\end{pmatrix}\Psi, 
\end{equation}
with 
$g$
\begin{equation}
g=\frac{1}{\sqrt{1-{x_{2k+1}}^2}} (x_{2k}\bold{1}_{2^{k-1}}+ix_i\gamma_i), 
\end{equation}
$A$ and $F$  are transformed as 
\begin{align}
&A~\rightarrow~g^{\dagger}Ag -ig^{\dagger}dg,\nn\\
&F~\rightarrow~g^{\dagger}Fg. \label{so2ktranssforgaugef}
\end{align}
The homotopy theorem guarantees the non-trivial bundle topology of the $SO(2k)$ monopole on $S^{2k}$\footnote{For $k=2, 4$ we have two $\mathbb{Z}$s: $\pi_3(SO(4))\simeq \mathbb{Z}\oplus \mathbb{Z}$ and  $\pi_7(SO(8))\simeq \mathbb{Z}\oplus \mathbb{Z}$.}:
\be
\pi_{2k-1}(SO(2k))\simeq \mathbb{Z}, 
\ee
which is measured by the $k$th Chern-number: 
\begin{equation}
c_k=\frac{1}{k!(2\pi)^k} \int_{S^{2k}}\text{tr}F^k. 
\label{chernandfk}
\end{equation}
In low dimensions, (\ref{chernandfk}) yields 
\begin{align}
&c_{k=1}=\frac{1}{2\pi}\int_{S^2}\text{tr}F,\nn\\
&c_{k=2}=\frac{1}{8\pi^2}\int_{S^4}\text{tr}F^2,\nn\\
&c_{k=3}=\frac{1}{48\pi^3}\int_{S^6}\text{tr}F^3,\nn\\
&c_{k=4}=\frac{1}{384\pi^4}\int_{S^8}\text{tr}F^4.
\end{align}
For the $SO(2k)$ fully symmetric representation $\overbrace{[\frac{I}{2},\frac{I}{2},\cdots,\frac{I}{2}]}^{k}$,  the Chern-numbers are calculated as \cite{Kimura2004}
\begin{align}
&c_{k=1}=I,\nn\\
&c_{k=2}=\frac{1}{6}I(I+1)(I+2),\nn\\
&c_{k=3}=\frac{1}{360}(I+1)(I+2)^2(I+3)(I+4),\nn\\
&c_{k=4}=\frac{1}{302400} I(I+1)(I+2)^2(I+3)^2(I+4)^2(I+5)(I+6), 
\label{chernnumberslowdim}
\end{align}
which correspond to the monopole charge or the number of magnetic fluxes on spheres.

\subsection{Non-commutative geometry in the lowest Landau level}\label{subsec:non-commutativegeo}

\begin{figure}[tbph]
\hspace{1cm}
\includegraphics*[width=150mm]{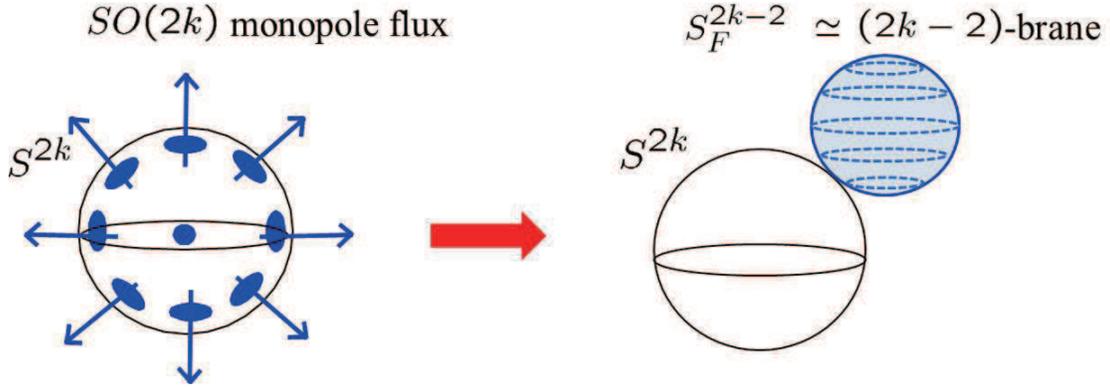}
\caption{The internal geometry of the $SO(2k)$ non-abelian flux is equivalent to the fuzzy-fibre  $S_F^{2k-2}$, and the $S_F^{2k-2}$  corresponds to $(2k-2)$-brane in the enlarged $(4k-2)$ dimensional  space.  }
\label{so2kflux}
\end{figure}

Following to the similar step in Sec.\ref{subsec:hopflag}, we can find how higher dimensional fuzzy sphere geometry emerges in the lowest Landau level. It should be noted since the monopole gauge field is non-abelian, and then the particle on $S^{2k}$ carries the $SO(2k)$ color degrees of freedom  like a ``quark''. 
The Lagrangian 
is given by 
\be
L=\frac{M}{2}\dot{x}_a\dot{x}_a-\dot{x}_a A_a, 
\label{lagrangeso2k}
\ee
where $x_ax_a=r^2$.  
In the lowest Landau level, the Lagrangian is reduced to 
\be
L=i\Psi^{\dagger}\frac{d}{dt}\Psi, 
\ee
with $\Psi$ (\ref{geneHopf}). By imposing the canonical quantization condition on $\Psi$ and $\Psi^{*}$, 
$x_a$ 
(\ref{genelahopf}) are effectively represented by  the  operators  
\be
X_a =\frac{\alpha}{2} \Psi^t \Gamma_a \frac{\partial}{\partial \Psi},  
\ee
which satisfy 
\be
[X_a, X_b]=i\alpha X_{ab}, 
\label{xaxabrelation}
\ee
where 
\be
X_{ab}=\alpha \Psi^t \Sigma_{ab}\frac{\partial}{\partial \Psi}, 
\ee
with $\Sigma_{ab}=-\frac{i}{4}[\Gamma_a, \Gamma_b]$.  
$X_a$ and $X_{ab}$ amount to $(2k+1)+k(2k+1)=(k+1)(2k+1)$ generators of  the $SO(2k+2)$ algebra, and 
$X_{ab}$ bring the ``extra'' degrees of freedom of fuzzy fibre $S_F^{2k-2}$ over $S^{2k}$. 
  It should be noted that the  coordinates of the external space  and those of the internal space are related by (\ref{xaxabrelation}) and they are same size matrices of the  $SO(2k+2)$ generators. Since they are similarly treated in the fuzzy algebra, there is no reason to distinguish the external  and  internal spaces in the lowest Landau level. It may be more natural to consider an enlarged  space that includes both external and internal spaces. 
Since the fuzzy-fibre coordinates $X_{ab}$ are the $SO(2k+1)$ generators, $X_{ab}$ can be represented as   
\be
X_{ab}=\alpha L_{ab}.   
\ee
Meanwhile   $L_{ab}\sim r^2 F_{ab}$ in the lowest Landau level (see Sec.\ref{sec:tenformononambuncg}). From these relations, we have 
\be
X_{ab}\sim \alpha r^2 F_{ab},
\ee
 which suggests the non-abelian field strength is equivalent to the fuzzy-fibre [see Fig.\ref{so2kflux}]. This identification coincides with the intuitive picture that the fuzzy-fibre realizes as the non-abelian flux of the monopole.  
In the 2D quantum Hall liquid, the $U(1)$ magnetic flux penetration induces a charged  excitation at the 
point where the flux is pierced. Similarly in higher dimensional quantum Hall liquid, the non-abelian flux penetration induces a point-like excitation on $S^{2k}$. Though the excitation is  ``point'' like  on  $S^{2k}$, the non-abelain flux matrix accommodates the  $S_F^{2k-2}$ geometry as its internal structure.   Remember that there is no distinction between the external and internal spaces in the lowest Landau level,  and so the ``internal'' space $S_F^{2k-2}$ can be regarded as an extended $(2k-2)$ dimensional  object, $(2k-2)$-brane, in the enlarged  $(4k-2)$ dimensional space. In this sense, the non-abelian flux penetration  induces $(2k-2)$-brane like excitation.      

\subsection{The $SO(2k+1)$ Landau model }\label{sec:so2k+1landaumodel}

In $d$-dimensional space, one-particle Hamiltonian under the influence of  gauge field is given by 
\be
H=-\frac{1}{2M}\sum_{a=1}^d {D_a}^2= -\frac{1}{2M}r^{1-d}\frac{\partial}{{\partial r}}r^{d-1}\frac{\partial}{{\partial r}}+\frac{1}{2Mr^2}\sum_{a<b}{\Lambda_{ab}}^2, 
\label{LandauHamilgeneric}
\ee
with 
$D_a=\partial_a+iA_a$ and  $\Lambda_{ab}=-ix_aD_b+ix_bD_a$ $(a,b=1,2,\cdots,d)$. $\Lambda_{ab} $  satisfy    
\begin{equation}
[\Lambda_{ab},\Lambda_{cd}]= i(\delta_{ac}\Lambda_{bd}+\delta_{bd}\Lambda_{ac}-\delta_{bc}\Lambda_{ad}-\delta_{ad}\Lambda_{bc})
-i(x_a x_c F_{bd}+x_{b}x_{d}F_{ac}-x_{b}x_{c}F_{ad}-x_{a}x_{d}F_{bc} ),  
\end{equation}
where $F_{ab}$ are the components of the field strength,  
$F_{ab}=-i[D_a, D_b]=\partial_a A_b- \partial_b A_a +i[A_a, A_b]$. 
Since the $SO(2k)$ non-abelian monopole (\ref{monopolegaugefield}) is located at the center of  $d=2k+1$  
dimensional space, its field strength is radially distributed and  the system respects the $SO(2k+1)$ rotational symmetry. We can construct the conserved $SO(2k+1)$ angular momentum as  
\begin{equation}
L_{ab}=\Lambda_{ab}+r^2F_{ab}. 
\label{totalangulamoment}
\end{equation}
It is straightforward to verify that $L_{ab}$ act as the $SO(2k+1)$ generators: 
\begin{equation}
[L_{ab}, M_{cd}]= i(\delta_{ac}M_{bd}+\delta_{bd}M_{ac}-\delta_{bc}M_{ad}-\delta_{ad}M_{bc}),  
\end{equation}
where $M_{ab}=L_{ab}, \Lambda_{ab}, F_{ab}$.  
For a particle on $2k$-sphere,  (\ref{LandauHamilgeneric}) is reduced to the $SO(2k+1)$ Landau Hamiltonian: 
\begin{equation}
H=\frac{1}{2Mr^2} \sum_{a<b}{\Lambda_{ab}}^2.  
\label{so2kL1LLham}
\end{equation}
Due to the existence of the $SO(2k+1)$ symmetry, one may readily derive the eigenvalues of (\ref{so2kL1LLham}) by a group theoretical method.   
With the orthogonality $\Lambda_{ab}F_{ab}=F_{ab}\Lambda_{ab}$, (\ref{so2kL1LLham}) is rewritten as 
\be
H=\frac{1}{2Mr^2}(\sum_{a<b}{L_{ab}}^2-\sum_{a<b}{F_{ab}}^2)=\frac{1}{2Mr^2}(\sum_{a<b} {L_{ab}}^2-\sum_{\mu<\nu}{\Sigma^{+}_{\mu\nu}}^2),
\ee
where $\sum_{a<b}{F_{ab}}^2=\sum_{\mu<\nu}{\Sigma_{\mu\nu}^{\pm}}^2$ was used. 
We adopt the fully symmetric representation 
\be 
({I}/{2})\equiv \overbrace{[\frac{I}{2},\frac{I}{2}, \cdots,\frac{I}{2}]}^k 
\ee
for the $SO(2k)$ Casimir $\sum_{\mu<\nu} {\Sigma^+_{\mu\nu}}^2$ , and  
 the irreducible representation   
\be
(n,I/2)\equiv \overbrace{[n+\frac{I}{2},\frac{I}{2},\frac{I}{2}, \cdots,\frac{I}{2}]}^k  
\ee
 for the $SO(2k+1)$ Casimir $\sum_{a<b}{L_{ab}}^2$ ($n$ denotes the Landau level index), and  
 then the energy eigenvalues are derived as\footnote{
In the thermodynamic limit, $r,I\rightarrow \infty$ with $I/r^2$ fixed, the energy eigenvalues (\ref{energyeigenso2k+1}) are reduced to 
\begin{equation}
E_n\rightarrow \frac{I}{2Mr^2}(n+\frac{1}{2}k). 
\end{equation}
The lowest Landau level energy, $E_{LLL}=\frac{I}{4Mr^2}k$, is equal to  $k$ times  the lowest Landau level energy of the 2D (planar) Landau model, $\frac{B}{2M}=\frac{I}{4Mr^2}$. This is because that in the thermodynamic limit, the $2k$D fuzzy sphere is reduced to $k$ copies of  $2$D non-commutative plane.}     
\be
E_n=\frac{1}{2Mr^2}(C_{2k+1}(n,{I}/{2})-C_{2k}({I}/{2}))=\frac{1}{2Mr^2} \biggl(n(n+2k-1)+I(n+\frac{1}{2}k)\biggr), 
\label{energyeigenso2k+1}
\ee
where $C_{2k+1}(n,{I}/{2})$ and $C_{2k}({I}/{2})$ respectively represent  
the $SO(2k+1)$ and $SO(2k)$ Casimir eigenvalues for $(n,I/2)$ and $(I/2)$:  
\begin{subequations}
\begin{align}
&C_{2k+1}(n,{I}/{2})=n^2+n(I+2k-1)+\frac{1}{4}Ik(I+2k), 
\label{so2k+1casimireigenvaluessym} \\
&C_{2k}({I}/{2})=\sum_{\mu < \nu} {\Sigma^{\pm}_{\mu\nu}}^2 =\frac{1}{4}Ik(I+2k-2). 
\label{so2kcasimireigenvaluessym}
\end{align}
\end{subequations}
The degeneracy in the $n$th Landau level is given by 
\be
D_{n}(k,I)=\frac{2n+I+2k-1}{(2k-1)!!}\frac{(n+k-1)!}{n!(k-1)!}\frac{(I+2k-3)!!}{(I-1)!!}\cdot \frac{(n+I+2k-2)!}{(n+I+k-1)!}\prod_{l=1}^{k-2} \frac{(I+2l)}{(I+l)!}\prod_{l=1}^{k-1}\frac{l!}{(2l)!}.  
\ee
In particular for the lowest Landau level $(n=0)$, the representation is reduced to the $SO(2k+1)$ fully symmetric spinor repr. $({I}/{2})$,  and the degeneracy becomes to 
\be
D_{LLL}(k,I)=\prod_{l=1}^{k}\prod_{i=1}^l \frac{I+l+i-1}{l+i-1}.
\label{iterativedegeneracylll}
\ee
In low dimensions, 
\begin{align}
&k=1~:~D_{LLL}(1,I)=I+1,\nn\\
&k=2~:~D_{LLL}(2,I)=\frac{1}{6}(I+1)(I+2)(I+3),\nn\\
&k=3~:~D_{LLL}(3,I)=\frac{1}{360}(I+1)(I+2)(I+3)^2(I+4)(I+5),\nn\\
&k=4~:~D_{LLL}(4,I)=\frac{1}{302400}(I+1)(I+2)(I+3)^2(I+4)^2(I+5)^2(I+6)(I+7).  
\label{numberstatesllllowdim}
\end{align}
One may notice that  the lowest Landau level degeneracy (\ref{numberstatesllllowdim}) and the Chern number (\ref{chernnumberslowdim}) are related by the following simple formula:  
\be
c_k(I)=D_{LLL}(k, I-1). 
\label{relationcd}
\ee
This relation is indeed guaranteed by the index theorem  for $\it{arbitrary}$ $k$ [see Sec.\ref{sec:indextheorem}].

\subsection{The $SO(2k+1)$ spinor Landau model and index theorem}\label{sec:indextheorem}

Here, we consider a spinor particle on $S^{2k}$ in the $SO(2k)$ monopole background. 
The spinor particle carries the $SO(2k+1)$ spin degrees of freedom  coupled to the external $SO(2k)$ magnetic field through Zeeman term. We analyze the $SO(2k+1)$ spinor Landau problem with use of the formulation explored by Dolan \cite{Dolan2003}. 

In the presence the gauge field,  the Dirac operator on $d$-dimensional curved manifold is generally given by 
\be
\fsl{D}=\gamma^{\alpha}D_{\alpha}={e^{\alpha}}_{\mu}\gamma^{\mu}(\partial_{\alpha}+i\omega_{\alpha}+i\mathcal{A}_{\alpha}), 
\ee
where $\alpha$ stand for the intrinsic coordinates of the manifold,   $\omega_{\alpha}$ denote the spin connection of the manifold,  $\mu$ represent the coordinates of the $d$-dimensional  flat Euclidean space, and $\gamma^\mu$ are the $SO(d)$ gamma matrices: 
\be
\{\gamma^\mu, \gamma^{\nu}\}=2\delta^{\mu\nu}. ~~~(\mu,\nu=1,2\cdots,d)
\ee
For symmetric ($\equiv$ torsion free) manifold, the square of the Dirac operator is given by the following Lichnerowicz formula \cite{Lichnerowicz1963}:   
\be
(-i\fsl{D})^2=-\Delta+\mathcal{F}_{\alpha\beta}\otimes \sigma^{\alpha\beta}+\frac{\mathcal{R}}{4} 
\label{squareDexpand}
\ee
where the Laplacian $\Delta$ and the field strength $\mathcal{F}_{\alpha\beta}$ are respectively  given by 
\begin{align}
&\Delta=\frac{1}{\sqrt{g}}\nabla_{\alpha}(\sqrt{g}g^{\alpha\beta}\nabla_{\beta})=g^{\alpha\beta}(\nabla_{\alpha}\nabla_{\beta}-\Gamma^{\gamma}_{~~\alpha\beta}\nabla_{\gamma}), \nn\\ 
&\mathcal{F}_{\alpha\beta}=\partial_{\alpha}\mathcal{A}_{\beta}-\partial_{\beta}A_{\alpha}+i[\mathcal{A}_{\alpha},\mathcal{A}_{\beta}],  
\end{align}
 and  $\mathcal{R}$ denotes the scalar curvature.  The second term on the right-hand side of (\ref{squareDexpand}), 
$\sigma_{\alpha\beta}\mathcal{F}^{\alpha\beta}=e^a_{~\alpha}e^b_{~\beta}\sigma_{ab}\mathcal{F}_{\alpha\beta}$, represents the  Zeeman term.  As readily verified from the Lichnerowicz formula, in the absence of  the Zeeman term, the Dirac operator does not have zero-eigenvalues  on manifolds with positive scalar curvature,  since the eigenvalues of Laplacian are semi-positive definite. Meanwhile in the presence of the gauge field strength, the Zeeman term may cancel the contribution from the curvature term to give zero-eigenvalues for $(-i\fsl{D})^2$.  This cancellation indeed occurs in the present case, and the zero-modes of the Dirac operator  are identified with the lowest Landau level basis states whose spin direction is opposite to the external magnetic field.     
When the gauge group is identical to the holonomy group of the coset $\mathcal{M}\simeq G/H$, (\ref{squareDexpand}) can be expressed by the group theoretical quantities\cite{Dolan2003}: 
\be
(-i\fsl{D})^2=C(G)-C(H, {R})+\frac{\mathcal{R}}{8},  
\label{squarediracoperator}
\ee
where $C(G)$ represents (quadratic) Casimir for the isometry group $G$ and $C(H, R)$ denotes  (quadratic) Casimir for the holonomy group $H$ made by the gauge group representation ${R}$.   
With (\ref{squarediracoperator}) we are able to derive the eigenvalues of $(-i\fsl{D})^2$ by using a simple group theoretical method. 

For  $S^{2k}\simeq SO(2k+1)/SO(2k)$,  we propose the $SO(2k+1)$ spinor Landau Hamiltonian as 
\be
H=\frac{1}{2M}(-i\fsl{D})^2=\frac{1}{2M}(C_{2k+1}-C_{2k})+\frac{{1}}{8M}k(2k-1),  
\label{simpleformulaford2}
\ee
where we used  the Ricci scalar of $S^{2k}$\footnote{The $SO(2k)$ Casimir for the  fundamental representation (\ref{so2kcasimireigenvaluessym})  ($I=1$) is equal to the Ricci scalar of $S^{2k}$; $\sum_{\mu <\nu}{\sigma_{\mu\nu}}^2=\frac{k}{4}(2k-1)=\frac{1}{8}\mathcal{R}$.} 
\be
\mathcal{R}=2k(2k-1).  
\ee
For the irreducible representations
\bse
\begin{align}
&(n, J)\equiv [\overbrace{n+J,J,\cdots,J}^{k}], ~~~~\text{for}~~ SO(2k+1)\\
&(\frac{I}{2})\equiv [\overbrace{\frac{I}{2},\frac{I}{2},\cdots,\frac{I}{2}}^{k}]~~~~~~~~~~~~~\text{for}~~ SO(2k), 
\end{align}
\ese
the Casimir eigenvalues are respectively given by 
\bse
\begin{align}
&C_{2k+1}(n,J)=n^2+n(2J+2k-1)+kJ(J+k),\\
&C_{2k}(\frac{I}{2})=k\frac{I}{2}(\frac{I}{2}+k-1), 
\end{align}
\ese
and the eigenvalues of (\ref{simpleformulaford2}) are derived as 
\be
E(n,J)=\frac{1}{2M}(n^2+n(2J+2k-1)+k(J(J+k) -\frac{I}{2}(\frac{I}{2}+k-1)))+\frac{1}{8M}k(2k-1), 
\label{levelnonrelaso2k+1}
\ee
and the  $n$th Landau level degeneracy is obtained as 
\begin{align}
D_n(k,2J)&=\frac{2n+2J+2k-1}{(2k-1)!!}\frac{(n+k-1)!}{n!(k-1)!}\nn\\
&\cdot\prod_{i=1}^{k-1}(2J+2i-1)\cdot \prod_{i=2}^k\frac{n+2J+2k-i}{2k-i}
\cdot \prod_{l=1}^{k-2}\prod_{i=l+2}^k \frac{2J+2k-i-l}{2k-i-l}.
\label{degeneracyof2jrela}
\end{align}
For the spinor particle\footnote{For the scalar particle, we substitute 
\be
J=\frac{I}{2} 
\ee
to (\ref{simpleformulaford2}) to derive the energy eigenvalues (\ref{energyeigenso2k+1}): 
\be
H-\frac{1}{8M}k(2k-1)=\frac{1}{2M}(C_{2k+1}(n,J)-C_{2k}({I}/{2}))|_{J=\frac{I}{2}}=\frac{1}{2M}(n^2+n(I+2k-1)+\frac{1}{2}Ik). 
\ee
}, we take 
\be
J=\frac{I}{2}\pm \frac{1}{2}, 
\ee
where for $+$ ($\uparrow$ spin state),  $I\ge 0 $, while  for $-$ ($\downarrow$ spin state), $I\ge 1$. 
This implies that the spin polarization due to the Zeeman effect effectively changes the strength of magnetic flux by $\pm\frac{1}{2}$ according to the direction of spin.  
In accordance with $\pm$ sector,  (\ref{levelnonrelaso2k+1}) is block diagonalized as  
\be
\begin{pmatrix}
E_+(n) & 0 \\
0 & E_-(n)
\end{pmatrix}, 
\ee
where $E_{\pm}(n) \equiv E(n, J)_{J=\frac{I}{2}\pm \frac{1}{2}}$: 
\begin{align}
&E_+(n)=\frac{1}{2M}(n^2+n(I+2k)+k(I+k)),\nn\\
&E_-(n)=\frac{1}{2M}(n^2+n(I+2k-2)),  \label{eigenpmress}
\end{align}
whose  degeneracies are respectively given by $D_{n}(k,I+ 1)$ and $D_n(k, I-1)$ through the formula (\ref{degeneracyof2jrela})\footnote{
It can be confirmed that $E_+(n)|_{I=0}$ (\ref{eigenpmress}) and $D_n(k,2J=I+1)|_{I=0}=D_n(k,1)$ (\ref{degeneracyof2jrela}) respectively reproduce  the eigenvalues and the degeneracy of the free Dirac operator without gauge field \cite{Sulanke1978,Trautman1995,CamporesiHiguchi1995,Balachandranetal2002}:  
\begin{align}
&{\sqrt{2M E_+(n)}}|_{I=0}=n+k, \nn\\
&D_n(k,1)=2^k 
\begin{pmatrix}
n+2k-1 \\
n
\end{pmatrix}.  
\end{align}
}. 
In low dimensions, (\ref{eigenpmress}) reads as 
\begin{align}
&S^2~:~~\begin{pmatrix}
E_+(n) & 0 \\
0 & E_-(n)
\end{pmatrix}\biggl|_{k=1}= \frac{1}{2M} \begin{pmatrix}
(n+1)(n+ I+1) & 0 \\
0 & n(n+I)
\end{pmatrix}
,\nn \\
&S^4~:~~\begin{pmatrix}
E_+(n) & 0 \\
0 & E_-(n)
\end{pmatrix}\biggl|_{k=2}= \frac{1}{2M}\begin{pmatrix}
n^2+n(I+4)+2(I+2) & 0 \\
0 & n^2+n(I+2)
\end{pmatrix}, \nn\\
&S^6~:~~\begin{pmatrix}
E_+(n) & 0 \\
0 & E_-(n)
\end{pmatrix}\biggl|_{k=3}=\frac{1}{2M}\begin{pmatrix}
n^2+n(I+6)+3(I+3) & 0 \\
0 & n^2+n(I+4)
\end{pmatrix}, \nn\\
&S^8~:~~\begin{pmatrix}
E_+(n) & 0 \\
0 & E_-(n)
\end{pmatrix}\biggl|_{k=4}=\frac{1}{2M}
\begin{pmatrix}
n^2+n(I+8)+4(I+4) & 0 \\
0 & n^2+n(I+6)
\end{pmatrix}. 
\end{align}
The Landau level energy spectrum is bounded by zero for the  lowest Landau level basis states ($n=0$) with $\downarrow$ spin: 
\be
E_-(n=0)=0, 
\ee
and the number of  the zero-energy states is  given by 
\be
D_{LLL}(k, I-1). 
\ee
Since the Hamiltonian is the square of the Dirac operator, 
 the zero-energy eigenstates correspond to the zero-modes of the Dirac operator: 
\be
\text{Ind}(i\fsl{D})=D_{LLL}(k, I-1). 
\ee
The index theorem tells that 
the number of zero-modes is equal to the topological charge of the non-trivial gauge configuration: 
\be
\text{Ind}(i\fsl{D})=c_k.  
\ee
In the present case,  $c_k$ denotes the $k$th Chern number of the $SO(2k)$ monopole (\ref{chernandfk}). 
We thus verified (\ref{relationcd}) for arbitrary $k$.

\subsection{Laughlin-like wavefunction}

 For higher dimensional quantum Hall effect, the particles carry the $SO(2k)$ color degrees of freedom with the geometry $S_F^{2k-2}$, and the total space will be given by 
\be 
(\boldsymbol{x}, \boldsymbol{y})~\in~ S^{2k}\times S^{2k-2}, 
\ee
where $\boldsymbol{x}=(x_1,x_2,\cdots,x_{2k+1})$ with $\sum_{a=1}^{2k+1}x_ax_a=r^2$ denotes the basemanifold $S^{2k}$ while $\boldsymbol{y}=(y_1, y_2, \cdots, y_{2k-2})$ with $\sum_{i=1}^{2k-1}y_iy_i=r^2$ represents the coordinates on $(2k-2)$-dimensional internal space $S^{2k-2}$ (which is regarded as the classical counterpart of fuzzy bundle coordinates $X_i$ (\ref{expanstionxmunusxi})).    
The coordinates of the total space $S^{2k}\otimes S^{2k-2}$ is represented by  
\be
\Psi(\boldsymbol{x})=\frac{1}{\sqrt{2r(r+x_{2k+1})}}
\begin{pmatrix}
(r+x_{2k+1}) \psi\\
(x_{2k}+i\gamma_i x_i) \psi
\end{pmatrix},  
\ee
where $\psi$ denotes $2^{k-1}$ component spinor giving  the internal coordinates by the relation:   
\be
\psi^{\dagger}\gamma_i\psi=y_i. 
\ee
The lowest Landau level basis states can be constructed  by taking a fully symmetric product of the components of $\Psi(\boldsymbol{x})$ : 
\be
\Psi_{m_1,m_2,\cdots,m_{2k}}(\boldsymbol{x})=\frac{1}{\sqrt{m_1! m_2 ! \cdots m_{2k}!}} \Psi_{1}^{m_1}(\boldsymbol{x})\Psi_{2}^{m_2}(\boldsymbol{x})\cdots \Psi_{2k}^{m_{2k}}(\boldsymbol{x}), 
\ee
 with $m_1+m_2+\cdots+m_{2k}=I$.  
 For $m=1$ the particles occupy all the lowest Landau level states on $S^{2k}$, and so the total particle number $N$ is  given by 
\be
N\equiv d(k,I)\equiv  \frac{D(k,I)}{D(k-1,I)}=\frac{(k-1)!}{(2k-1)!}\frac{(I+2k-1)!}{(I+k-1)!}
~\sim~I^k, 
\label{NandIkrelation}
\ee
where $D(k, I)$ denotes the number of states of the total space $S_F^{2k}$, and $D(k-1, I)$ stands for the number of states of  the fuzzy-fibre $S_{F}^{2k-2}$.  
For $I/2~\rightarrow~mI/2$, the state number on $S^{2k}$ changes as 
\be 
d(k, mI)=\frac{D(k,mI)}{D(k-1,mI)}=\frac{(k-1)!}{(2k-1)!}\frac{(mI+2k-1)!}{(mI+k-1)!}~\sim~(mI)^k. 
\ee
With use of the Slater determinant, 
the  Laughlin-like groundstate wavefunction  is constructed as  
\be
\Psi_{\text{Lin}}(\boldsymbol{x}_1,\boldsymbol{x}_2,\cdots ,\boldsymbol{x}_N)=
(\epsilon_{A_1 A_2 \cdots A_N} \Psi_{A_1}(\boldsymbol{x}_1)\Psi_{A_2}(\boldsymbol{x}_2)\cdots \Psi_{A_N}(\boldsymbol{x}_N))^{m}, 
\ee
where $A=(m_1,m_2,\cdots,m_{2k})$ and $m$ is taken as an odd integer to keep the Fermi statistics of the particles. When the power of $\Psi_A$ changes from $1$ to $m$, the monopole charge changes from $I$ to $mI$, and then $\Psi_{\text{Lin}}$  corresponds to the groundstate of $2k$D quantum Hall liquid at the filling factor: 
\be
\nu_{2k}=\frac{N}{d(k, mI)}\simeq \frac{1}{m^k}. 
\label{fillingmembr}
\ee
Notice that since $m$ is an odd inter, $\nu_{2k}$ is also the inverse of an odd integer.  
From the perspective of the original basemanifold $S^{2k}$,  $\Psi_{\text{Llin}}$ denotes the incompressible liquid made of the particles. However, from the emergent $(4k-1)$D space-time point of view, the particle corresponds to $(2k-2)$-brane, and $\Psi_{\text{Llin}}$ is alternatively  interpreted as a many-body state of membranes.

\section{Tensor Monopole Fields from Non-Abelian Monopole Fields}\label{sec:tensormono}

We discussed the non-abelian monopoles whose gauge group is compatible with the holonomy of sphere. In this section, we introduce another type of monopole, the tensor monopole \cite{Nepomechie1985, Teitelboim1986}  whose gauge group is  $U(1)$ and gauge field is an antisymmetric tensor\footnote{The antisymmetric tensor gauge field is realized as a solution of the Kalb-Ramond equation and also referred to as the Kalb-Ramond field \cite{KalbRamond1974}.}.

\subsection{Tensor monopole fields}

To begin with, we review several basic properties of $n$-form tensor gauge field \cite{Teitelboim1986}:
\begin{equation}
C_n=\frac{1}{n!}C_{a_1a_2\cdots a_n}dx_{a_1}dx_{a_2}\cdots dx_{a_n}
\end{equation}
where $C_{a_1 a_2\cdots a_n}$ represent a totally antisymmetric tensor gauge field. Notice that $C_{a_1a_2\cdots a_n}$ is $\it{not}$ a matrix-valued gauge field but a tensor extension of the $U(1)$ gauge field.    
Like the ordinary $U(1)$ gauge theory, the field strength is defined as 
\begin{equation}
G_{n+1}=dC_{n}=\frac{1}{(n+1)!}G_{a_1a_2\cdots a_{n+1}}dx_{a_1}dx_{ a_2}\cdots dx_{ a_{n+1}},
\end{equation}
where 
\begin{equation}
G_{ a_1  a_2\cdots   a_{n+1}}= \frac{1}{n!}\partial_{[ a_1}C_{ a_2\cdots  a_{n+1}]}.
\end{equation}
For instance, 
\begin{align}
&n=2~:~G_{ a b c}=\partial_{ a} C_{ b c}+\partial_{ b} C_{ c a}+\partial_{ c} C_{ a b},  \nn\\
&n=3~:~G_{ a b c d}=\partial_{ a} C_{ b c d}-\partial_{ b} C_{ c d a}+\partial_{ c} C_{ d a b}-\partial_{ d} C_{ a b d}. 
\label{fromctog3}
\end{align}
The $U(1)$ gauge symmetry  is incorporated in the following way.  The $U(1)$ gauge transformation is given by 
\begin{equation}
C_{n}~\rightarrow ~C_{n}+d\Lambda_{n-1}, 
\label{cdlambdagauge}
\end{equation}
with 
\begin{equation}
\Lambda_{n-1}=\frac{1}{(n-1)!}\Lambda_{ a_1 a_2\cdots  a_{n-1}}dx_{ a_1}dx_{ a_2}\cdots dx_{ a_{n-1}}.
\end{equation}
It is obvious that the field strength $G$ is invariant under (\ref{cdlambdagauge}). In terms of  the tenor components, the gauge transformation is represented as  
\begin{equation}
C_{ a_1 a_2\cdots  a_n}~\rightarrow~C_{ a_1 a_2\cdots  a_n}+\frac{1}{(n-1)!}\partial_{[ a_1}\Lambda_{ a_2\cdots  a_n]}.
\end{equation}
For instance, 
\begin{align}
&n=2~:~C_{ a b}~\rightarrow~C_{ a b}+\partial_{ a} \Lambda_{ b}-\partial_{ b} \Lambda_{ a}, 
 \nn\\
&n=3~:~C_{ a b c}~\rightarrow~C_{ a b c}+\partial_{ a} \Lambda_{ b c}+\partial_{ b} \Lambda_{ c a}+\partial_{ c} \Lambda_{ a b}.  
\label{ctrans3}
\end{align}
It is a simple exercise to see that (\ref{fromctog3}) is invariant under  (\ref{ctrans3}). 
The field strength of the $U(1)$ tensor monopole located at  the origin of $(n+2)$D Euclidean space is given by  
\be
G_{a_1 a_2 \cdots a_{n+1}}=g\frac{1}{ r^{n+2}} \epsilon_{a_1 a_2 \cdots a_{n+2}}x_{a_{n+2}}, 
\label{nrankgaugestrength}
\ee
where $g$ denotes the charge of $U(1)$ tensor monopole. The integral of the gauge field strength over $S^{n}$ yields 
\be
\int_{S^{n+1}} G_{n+1}=g\mathcal{A}(S^{n+1}), 
\ee
where $\mathcal{A}(S^{n+1})$ represents the area of $S^{n+1}$.

\subsection{Correspondence between field strengths of  monopoles}\label{subsec:relationsfieldmoono}

The non-abelian and tensor monopoles are two different extensions of the Dirac monopole in terms of  internal and external indices.    As discussed in Sec.\ref{sec:higherdimNCG}, there is no reasonable distinction between the external and internal spaces in the lowest Landau level, and so it is expected that non-abelian and tensor monopoles should be ``equivalent'' in some sense. 
Interestingly,  for the $SU(2)$ monopole and 3-rank tensor monopole, their connection has  already been  pointed out, at least for fundamental representation (quaternions) \cite{WuZee1988} and for the integral form \cite{DemlerZhang1998}.  As a natural generalization of  these  results,  we establish  connection between  tensor  and non-abelian monopoles for fully symmetric representation in arbitrary even dimension. 
 In the following, we take $n$ as an odd integer, $n=2k-1$ and the monopole at the center of $S^{2k}$ [Table \ref{table:correspDNYM}].  The tensor monopole gauge field (\ref{nrankgaugestrength}) takes the following form:  
\be 
G_{a_1 a_2 \cdots a_{2k}}=g_k\frac{1}{ r^{2k+1}} \epsilon_{a_1 a_2 \cdots a_{2k+1}}x_{a_{2k+1}}. 
\label{formoftensorG} 
\ee 
\begin{table}
\begin{center}
   \begin{tabular}{|c|c|c|}\hline
    /      &   Non-abelian monopole &  Tensor  monopole \\ \hline
Sphere             &  $S^{2k}$      & $S^{2k}$                \\ \hline 
Gauge group              &  $SO(2k)$      &  $U(1)$               \\ \hline 
Rank of gauge field               &   1   &   $2k-1$              \\ \hline   
Rank of field strength              &  2     &  $2k$               \\ \hline  
    \end{tabular}       
\end{center}
\caption{ Relations between the non-abelian monopole and the tensor monopole. 
 }
\label{table:correspDNYM}
\end{table}
We fix the ratio  between two monopole charges, $c_k$ (\ref{chernandfk}) and $g_k$, by imposing the condition: 
\be
\int_{S^{2k}}G_{2k}=\text{tr} \int_{S^{2k}} {F}^{k}. 
\label{relfkandg}
\ee
From 
\be
\int_{S^{2k}}G_{2k}=g_k \mathcal{A}(S^{2k}) 
\ee
with 
\be
\mathcal{A}(S^{2k})=\frac{2^{k+1}\pi^{k}}{(2k-1)!!},  
\ee
the relation between two monopole charges is determined as 
\be
g_k=\frac{(2k)!}{2^{k+1}}c_k.  
\label{gkckrel}
\ee
Eq.(\ref{relfkandg}) is rather ``trivial'', 
 since we are always able to impose (\ref{relfkandg}) by fixing  the ratio between the two monopole charges. 
What we really need to verify is the $\it{local}$ non-abelian and tensor monopole relation: 
\be
G_{2k}=\text{tr}~{F}^{k}.  
\label{conditionlocal}
\ee
To prove (\ref{conditionlocal}) we take a brute force method:  
We substitute the explicit form of $F$ (\ref{nonabelstrength}) to the right-hand side of (\ref{conditionlocal}) to see whether we can derive $G$ (\ref{formoftensorG}) on the left-hand side  under the identification (\ref{gkckrel}).   
For the component relation between $G_{a_1a_2\cdots a_{2k}}$ ($a_1,a_2,\cdots, a_{2k}=1,2,\cdots, 2k+1$) and $F_{ab}$,  the local relation (\ref{conditionlocal}) can be rewritten as
\footnote{ 
Here, we used 
\begin{equation}
G_{2k}=\frac{1}{(2k)!}G_{a_1a_2\cdots a_{2k}}dx_{a_1}dx_{a_2}\cdots dx_{a_{2k}}.  
\end{equation}
and  
\begin{equation}
\tr{F^k}= \frac{1}{2^k}\tr(F_{a_1a_2}\cdots F_{a_{2k-1}a_{2k}})dx_{a_1}dx_{a_2}\cdots dx_{a_{2k}}.  
\end{equation}
} 
\begin{equation}
G_{a_1 a_2 \cdots a_{2k}}=\frac{1}{2^k}\epsilon_{a_1 a_2 \cdots a_{2k+1}}\epsilon_{b_{a_1} b_{a_2}\cdots b_{a_{2k}} a_{2k+1}}\tr (F_{b_{a_1}b_{a_2}}\cdots F_{b_{a_{2k-1}}b_{a_{2k}}}). 
\label{U1tensormonopolenonabefields}
\end{equation}
For instance, 
\begin{equation}
G_{1 2 \cdots {2k}}=\frac{1}{2^k}\epsilon_{{\mu_1} {\mu_2}\cdots {\mu_{2k}}}\tr (F_{{\mu_1}{\mu_2}}\cdots F_{{\mu_{2k-1}}{\mu_{2k}}}), 
\label{122kcompG}
\end{equation}
where $\mu_1, \mu_2, \cdots, \mu_{2k}=1,2,\cdots, 2k$. We 
substitute (\ref{nonabelstrength}) to the right-hand side of (\ref{122kcompG}) and perform a straightforward calculation with use of the formulae for the $SO(2k)$ matrices (\ref{formulageneso2kL1I=1}),   
and then we find the right-hand side of (\ref{122kcompG}) gives 
\be
G_{12\cdots 2k}=\frac{(2k)!}{2^{k+1}r^{2k+1}}x_{2k+1}. 
\label{g1232kex}  
\ee
In the covariant notation, (\ref{g1232kex}) is expressed as 
\be
G_{a_1a_2\cdots a_{2k}}=\frac{(2k)!}{2^{k+1}r^{2k+1}} \epsilon_{a_1a_2\cdots a_{2k+1}} x_{a_{2k+1}}
\label{spinorrepfieldstreu1}
\ee
or 
\be
G_{2k}
=\frac{1}{2^{k+1}r^{2k+1} }  \epsilon_{a_1a_2\cdots a_{2k+1}}{x_{a_{2k+1}}} dx_{a_1}dx_{a_2}\cdots dx_{a_{2k}}. 
\ee
For instance, 
\begin{align}
~~~~~U(1)~~:~~~~~~~~~~~G_{ij}
&=\frac{1}{2r^3}\epsilon_{ijk}x_k,~~~~~~~~~~~~~~~~~~~~~~~~~~~~~~~~~~~~~~~~~~~~~~~(i,j,k=1,2,3)\nn\\
~~~~~SU(2)~:~~~~~~~~G_{abcd} 
&=\frac{3}{r^5}\epsilon_{abcde}x_e   ,~~~~~~~~~~~~~~~~~~~~~~~~~~~~~~~~~~~(a,b,c,d,e=1,2,3,4,5)\nn\\   
~~~~~~SO(6)~:~~~G_{a_1a_2\cdots a_6}
&=\frac{45}{r^7}\epsilon_{a_1a_2\cdots  a_6 a_7}x_{a_7},~~~~~~~~~~~~~~~~~~~~~~~(a_1,a_2,\cdots, a_7=1,2,\cdots,7)\nn\\
~~~~~~SO(8)~:~G_{a_1a_2 a_3\cdots  a_8}
&=\frac{1260}{r^9}\epsilon_{a_1a_2 \cdots a_8  a_9}x_{a_9}.~~~~~~~~~~~~~~~~~~~~~(a_1,a_2,\cdots, a_9=1,2,\cdots,9)
\end{align}
We thus demonstrated the derivation of the tensor monopole gauge field $G$ from $\tr F^k$.  
Furthermore in terms of a general symmetric representation of the $SO(2k)$\footnote{$I=1$ corresponds to the spinor representation.}
\begin{equation}
({I}/{2})\equiv \overbrace{[\frac{I}{2},\frac{I}{2},\cdots,\frac{I}{2}]}^{k}, \nn
\end{equation}
we can derive a generic expression for the $U(1)$ tensor field strength as 
\begin{align}
G_{a_1 a_2\cdots a_{2k}}&=\frac{(2k)!I}{2^{k+2}}C(k, I)D(k-1,I) \frac{1}{r^{2k+1}}\epsilon_{a_1a_2\cdots a_{2k+1}} {x_{a_{2k+1}}}\nonumber\\
&=\frac{I}{2}C(k, I) D(k-1,I)G_{a_1 a_2 \cdots a_{2k}}^{(I=1)},
\label{symmuifiestrcom}
\end{align}
where $C(k, I)$ and $G_{a_1a_s\cdots a_{2k+1}}^{(I=1)}$ are respectively given by (\ref{defofckI}) and (\ref{spinorrepfieldstreu1}). 
Here, we used the  formulae for the symmetric representation  
(\ref{relationgeneralIso2k+1gene}). 
One can confirm the symmetric representation(\ref{symmuifiestrcom}) for $I=1$ reproduces  (\ref{spinorrepfieldstreu1}) by the formula 
\be
D_{LLL}(k,I=1)=\frac{(2k)!!}{k!}=2^k. \label{formula2n2}
\ee
With (\ref{formula2n2}) and the following formula about the lowest Landau level degeneracy   
\be
C(k, I) D_{LLL}(k-1, I) =\frac{(2k)!}{2^k I} D_{LLL}(k,I-1), 
\label{relationdk-1IdkI-1}
\ee
we finally find that $G$ takes an amazingly simple form\footnote{In differential form, (\ref{ggeneralck}) is represented as  
\be
G_{2k}=\frac{1}{2^{k+1}r^{2k+1}}c_k(I)\epsilon_{a_1a_2\cdots a_{2k+1}}x_{a_{2k+1}}dx_{a_1}dx_{a_2}\cdots dx_{a_{2k}}=c_k(I) G^{(I=1)}, 
\label{explicitu1tensorgen}
\ee
and hence the normalized $U(1)$ tensor monopole charge  
$q_k(I)\equiv \frac{1}{\int_{S^{2k}}G_{2k}^{(I=1)}}\int_{S^{2k}}G_{2k}$, 
is identical to the Chern number: 
\be
q_k(I)=c_k(I). 
\label{Ckckequiv}
\ee
}:   
\be
G_{a_1 a_2 \cdots a_{2k}}=c_k(I) \cdot G_{a_1 a_2 \cdots a_{2k}}^{(I=1)}, 
\label{ggeneralck}
\ee
where  $G^{(I=1)}$ is  given by (\ref{spinorrepfieldstreu1}) 
and the relation (\ref{relationcd}) was used. 
From (\ref{ggeneralck}), we can read off the tensor monopole charge as $g_k=\frac{(2k)!}{2^{k+1}}c_k(I)$, which is consistent with the result (\ref{gkckrel}). 
In low dimensions, we have 
\begin{align}
G_{ij}&=\frac{1}{2r^3}I\epsilon_{ijk}x_k
,\nn\\
G_{abcd}&=\frac{1}{2r^5}I(I+1)(I+2)\epsilon_{abcde}x_e
,\nn\\
G_{a_1a_2\cdots a_6}&=\frac{1}{8 r^7}I(I+1)(I+2)^2(I+3)(I+4)\epsilon_{a_1 a_2 \cdots a_7} x_{a_7}
,\nn\\
G_{a_1a_2a_3 \cdots a_8}&=\frac{1}{240 r^9}I(I+1)(I+2)^2(I+3)^2(I+4)^2(I+5)(I+6) \epsilon_{a_1 a_2 \cdots a_9} x_{a_9}.
\label{generalIfieldstu1}
\end{align}
Thus, we verified the local non-abelian and tensor monopole correspondence (\ref{conditionlocal}) for generic fully symmetric representation in arbitrary even dimension.   

\subsection{Correspondence between  gauge fields of monopoles}\label{subsec:relationfields}

For non-abelian gauge field, we have \cite{Alavarez-GaumeGinsparg1984}
\begin{equation}
\text{tr}(F^{k}) =d L_{\text{CS}}^{(2k-1)}[A], 
\label{defnof2n1cslag}
\end{equation}
where $L_{\text{CS}}^{(2k-1)}$ represents the Chern-Simons term 
\begin{equation}
L_{\text{CS}}^{(2k-1)}[A]= k\int_0^1 dt~ \text{tr}(A(tdA+it^2A^2)^{k-1}).  
\label{generalexcernsimonsterm}
\end{equation}
Meanwhile for the tensor monopole gauge field,  we have seen 
\be
G_{2k}=dC_{2k-1}. 
\ee
From the non-abelian and tensor monopole correspondence (\ref{conditionlocal}), it is obvious that the tensor monopole gauge field is identical to the non-abelian Chern-Simons term: 
\be
C_{2k-1}=\text{tr}(L_{\text{CS}}^{(2k-1)}[A]). 
\label{nonabeliantensorcorres}
\ee
For instance, 
\begin{align}
C_{1}&=\text{tr}A,\nn\\
C_{3}
&=\text{tr}( AdA +\frac{2}{3}i A^3)=\tr(AF-\frac{1}{3}iA^3), \nn\\
C_{5}
&=\text{tr}( A(dA)^2 +\frac{3}{2}iA^3dA-\frac{3}{5} A^5)=\tr(AF^2-\frac{1}{2}iA^3F-\frac{1}{10}A^5), \nn\\
C_{7}
&=\text{tr}( A(dA)^3 +\frac{8}{5}iA^3(dA)^2+\frac{4}{5}iA(AdA)^2 -2A^5dA-\frac{4}{7}iA^7)\nonumber\\
&=\tr(AF^3-\frac{2}{5}iA^3F^2-\frac{1}{5}iAFA^2F-\frac{1}{5}A^5F+\frac{1}{35}iA^7).\label{expliciexcs}
\end{align}
Notice that $\tr(A^3F^2)\neq \tr(AFA^2F)$, since $A$ and $F$ are  matrix-valued quantities  
 and are not commutative. 
For components of (\ref{expliciexcs}), we have 
\begin{align}
C_i&=\text{tr}A_i,
\nn\\
C_{ a b c}&=\tr(A_{[ a}\partial_{ b}A_{ c]}+\frac{2}{3}iA_{[ a}A_{ b}A_{ c]})=\frac{1}{2}\tr(A_{[ a}F_{ b c]}-\frac{2}{3}iA_{[ a}A_{ b}A_{ c ]}), 
\nn\\
C_{abcde}&=\frac{1}{4}\tr(A_{[a}F_{bc}F_{de]} 
-i A_{[a}A_{b}A_{c}F_{de]}-\frac{2}{5}A_{[a}A_{b}A_{c}A_{d}A_{e]}), 
\nn\\
C_{a_1 a_2\cdots a_7}
&=\frac{1}{8}\text{tr}(A_{[a_1} F_{a_2a_3}F_{a_4a_5}F_{a_6a_7]}-\frac{4}{5}i A_{[a_1}A_{a_2}A_{a_3}F_{a_4a_5}F_{a_6a_7]}
-\frac{2}{5}i A_{[a_1}F_{a_2a_3}A_{a_4}A_{a_5}F_{a_6a_7]}\nn\\
&-\frac{4}{5}A_{[a_1}A_{a_2}A_{a_3}A_{a_4}A_{a_5}F_{a_6 a_7]}+\frac{8}{35}iA_{[a_1}A_{a_2}A_{a_3}A_{a_4}A_{a_5}A_{a_6}A_{a_7]}).  
\label{correpondcsandcfield}
\end{align}

The $SO(2k)$ gauge transformation acts as the  $U(1)$ gauge transformation for $C_{2k-1}$. For instance $k=2$, the non-abelian ($SU(2)$) gauge transformation (\ref{so2ktranssforgaugef}) acts to $C_{3}$ as 
\begin{equation}
C_{3}~\rightarrow~C_{3}-id(\tr Adgg^{\dagger})+\frac{1}{3}\tr(g^{\dagger}dg)^3.
\label{transsu2cfield}
\end{equation}
The second term on the right-hand side is the total derivative.  The third term  satisfies\footnote{$\tr(\alpha^{2n})=0$ for any one-form $\alpha=dx_{ a}\alpha_{ a}$.} 
\begin{equation}
d(\tr(g^{\dagger}dg)^3)=-\tr(g^{\dagger}dg)^4=0,   
\end{equation}
 and is locally expressed as a total derivative (Poincar$\acute{\text{e}}$ Lemma). 
Consequently, (\ref{transsu2cfield}) can be rewritten in the following form  
\be
C_{3}~\rightarrow~C_{3}+d\Lambda_{2}.  
\ee
In general,  the $SO(2k)$ gauge transformation acts as  $U(1)$ gauge transformation to  tensor gauge field (see Appendix \ref{appen:nonandu1charges} for more details): 
\begin{equation}
C_{2k-1}~\rightarrow~C_{2k-1}+d\Lambda_{2k-2}. 
\label{u1transc2k-1}
\end{equation}

 For practical applications, 
 it is important to derive the explicit form of the tensor monopole gauge field.  
With use of the general formula (\ref{correpondcsandcfield}), 
 we derive the tensor monopole gauge field from the non-abelian monopole in low dimensions.  
We substitute the non-abelian monopole field (\ref{monopolegaugefield}) to the right-hand side of the formula (\ref{correpondcsandcfield}). After a long but straightforward  calculations using  trace formulae of gamma matrices,  we obtain the following expressions for spinor representation:  
\begin{align}
C_i&=-\frac{1}{2r(r+x_3)}\epsilon_{ij3 }x_j,
\nn\\
C_{ a b c}&=-\frac{1}{r^3}\biggl(\frac{1}{r+x_5}+\frac{r}{(r+x_5)^2} \biggr)\epsilon_{abcd 5}x_d, 
\nn\\
C_{abcde}&=-\frac{9}{r^5}\biggl(\frac{1}{r+x_7}+\frac{r}{(r+x_7)^2}+\frac{2}{3}\frac{r^2}{(r+x_7)^3}\biggr)\epsilon_{abcdef7} x_{f}, 
\nn\\
C_{a_1 a_2\cdots a_7}
&=-
\frac{180}{r^7}\biggl(\frac{1}{r+x_9}
+\frac{r}{(r+x_9)^2}+\frac{4}{5}\frac{r^2}{(r+x_9)^3}+\frac{2}{5}\frac{r^3}{(r+x_9)^4}\biggr)\epsilon_{a_1a_2\cdots a_8 9 }x_{a_8}. 
\label{explicittensorlow}
\end{align}
Notice that $(2k-1)$ rank tensor monopole gauge field exhibits $k$th power string-like singularity.   
Similarly for fully symmetric representation, we obtain
\begin{align}
C_i&=-\frac{I}{2r(r+x_3)}\epsilon_{ij3 }x_j,\nn\\
C_{ a b c}&=-\frac{1}{6r^3}I(I+1)(I+2)\biggl(\frac{1}{r+x_5}+\frac{r}{(r+x_5)^2}\biggr)\epsilon_{abcd5}x_d, \nn\\
C_{abcde}&=-\frac{1}{40r^5}I(I+1)(I+2)^2(I+3)(I+4)\biggl(\frac{1}{r+x_7}+\frac{r}{(r+x_7)^2}+\frac{2}{3}\frac{r^2}{(r+x_7)^3}\biggr)\epsilon_{abcdef7} x_{f}\nn\\
C_{a_1 a_2\cdots a_7}
&=-
\frac{1}{1680 r^7}I(I+1)(I+2)^2(I+3)^2(I+4)^2(I+5)(I+6)\nn\\
&~~~~~~~~~~~~~~~~\times \biggl(\frac{1}{r+x_9}
+\frac{r}{(r+x_9)^2}+\frac{4}{5}\frac{r^2}{(r+x_9)^3}+\frac{2}{5}\frac{r^3}{(r+x_9)^4}\biggr)\epsilon_{a_1a_2\cdots a_8 9}x_{a_8}.  \label{fullysymcexplis}
\end{align}
For $I=1$, (\ref{fullysymcexplis}) is reduced to (\ref{explicittensorlow}).   
One may also confirm that (\ref{fullysymcexplis}) indeed gives the field strength (\ref{generalIfieldstu1}) through the formula: 
\be
G_{a_1 a_2 \cdots a_{2k}}=\frac{1}{(2k-1)!}\partial_{[a_1} C_{a_2 \cdots a_{2k-1}]}. 
\label{gfromctensor}
\ee

\subsection{Quantum Nambu geometry via tensor monopole}\label{sec:tenformononambuncg}

In the lowest Landau level, the covariant angular momentum is quenched, and then we have the  identification:  
\be
L_{ab}=\Lambda_{ab}+r^2F_{ab}~\sim~r^2 F_{ab}. 
\label{labandfablll}
\ee
In 3D, two rank antisymmetric tensor is equivalent to vector, and the angular momentum is directly related to the coordinates of fuzzy two-sphere (\ref{xiliequiv}). 
However in higher dimensions, two rank antisymmetric tensor is no longer equivalent to vector and  the angular momentum does not seem to  apparently be related to the coordinates of fuzzy sphere.   
 As mentioned in Sec.\ref{subsec:twomonop}, the quantum Nambu bracket implies the existence of  tensor monopole and we have shown the non-abelian and tensor monopole correspondence (\ref{conditionlocal}) or 
\be
\frac{1}{r^{2k+1}} x_a =\frac{2}{(2k)!c_k}\epsilon_{a a_1 a_2 \cdots a_{2k}}\tr (F_{a_1 a_2 } \cdots F_{a_{2k-1} a_{2k}}).  
\label{xfromfs}
\ee
The identification  ({\ref{labandfablll}) suggests that  (\ref{xfromfs}) becomes to  
\be
X_a=\frac{I}{(2k)! c_k} ~{\alpha}~\epsilon_{a a_1 a_2 \cdots a_{2k}}(L_{a_1 a_2}L_{a_3 a_4} \cdots L_{a_{2k-1} a_{2k}})  
\label{xafromlas}
\ee
in the lowest Landau level, and 
the coordinates of higher dimensional sphere  are now regarded as  the operators. 
 Eq.(\ref{xafromlas}) is a natural generalization of  (\ref{angularfromxa}). 

Let us consider the algebra for $X_a$.  For this purpose, it is useful to adopt the analogy between the algebras of $X_a$ and the covariant derivatives $-iD_a$  \cite{Bernevig2003}. 
For $S_F^2$ case, the algebra of $X_i$ is given by 
\be
[X_i,X_j]=i \alpha \epsilon_{ijk}X_k, 
\ee
while the covariant derivative gives  
\be
[-iD_i,-iD_j]=-iF_{ij}=-i\frac{1}{\alpha r^2}\epsilon_{ijk}x_k.  
\ee
One may notice the analogy: 
\be
[X_i,X_j] ~\leftrightarrow~-(\alpha r)^2[-iD_i,-iD_j].  
\label{commuanaxd} 
\ee
This analogy can hold  in higher dimensions [see Sec.\ref{subsec:twomonop}], and for evaluation of the Nambu bracket for $X_a$ we utilize the following identification: 
\be
[X_{a_1}, X_{a_2}, \cdots, X_{a_{2k}}] ~\leftrightarrow ~\frac{1}{D_{LLL}(k-1,I)} (-(\alpha r)^2)^k [-iD_{a_1}, -iD_{a_2}, \cdots, -iD_{a_{2k}}].   
\label{correspondxsandfs} 
\ee
The right-hand side gives 
\begin{align}
&[-iD_{a_1}, -iD_{a_2},\cdots, -iD_{a_{2k}}]
\nn\\
&=\frac{1}{2^k}\epsilon_{a_1 a_2 \cdots a_{2k+1}}
\epsilon_{b_{a_1}b_{a_2} 
\cdots b_{a_{2k}} a_{2k+1}}
[-iD_{b_{a_1}}, -iD_{b_{a_2}}] [-iD_{b_{a_3}}, -iD_{b_{a_4}}]\cdots [-iD_{b_{a_{2k-1}}}, -iD_{b_{a_{2k}}}]\nn\\
&=(-i\frac{1}{2})^k \epsilon_{a_1 a_2 \cdots a_{2k+1}}
\epsilon_{b_{a_1}b_{a_2} 
\cdots b_{a_{2k}} a_{2k+1}} F_{b_{a_1} b_{a_2}}F_{b_{a_3} b_{a_4}}\cdots F_{b_{a_{2k-1}} b_{a_{2k}}}, 
\label{dnalgebratof}
\end{align}
and the trace is evaluated as 
\be
\tr [-iD_{a_1}, -iD_{a_2},\cdots, -iD_{a_{2k}}]=(-i)^k \frac{(2k)!}{2^
{k+1}} D_{LLL}(k, I-1) \cdot \epsilon_{a_1 a_2 \cdots a_{2k+1}}  \frac{1}{r^{2k+1}}x_{a_{2k+1}}. 
\ee
Due to the relation (\ref{relationdk-1IdkI-1}), we obtain 
\be
[X_{a_1}, X_{a_2}, \cdots, X_{a_{2k}}] =i^k C(k, I) \alpha^{2k-1} \epsilon_{a_1 a_2 \cdots a_{2k+1}}X_{a_{2k+1}}, 
\ee
which is exactly equal to the quantum Nambu algebra for fuzzy sphere (\ref{fuzzynbracksphere}).

\section{Flux Attachment and Tensor Chern-Simons Field Theory for Membranes}\label{sec:cstheorymemb} 

  Here we discuss physical properties of  A-class topological insulator based on Chern-Simons  tensor field theory. We will see exotic concepts in 2D quantum Hall effect are naturally generalized in higher dimensions:  
\begin{itemize}
\item Flux attachment and composite particles \cite{Girvin-M-1987, Read-1989,  
Zhang-1992}
\item Effective topological field theory \cite{Zhang-H-K-1988,Zhang-1992}
\item Fractional statistics of quasi-particle excitations \cite{ArovasSW1984} 
\item Haldane-Halperin hierarchy \cite{haldane1983, Halperin1984}
 \\
~~~~~\vdots 
\end{itemize}

\subsection{Basic observations}

 Before going to the details, we summarize basic observations about the relevant physical concepts and associated mathematics  in higher dimensions.  
 
\begin{itemize}
\item  $(2k-1)$ rank tensor gauge field and $(2k-2)$-brane 
\end{itemize}
The $(2k-1)$ rank gauge field is naturally coupled to the $(2k-1)$ rank current of $(2k-2)$-brane.  
The membrane degrees of freedom is $\it{automatically}$ incorporated in 
the geometry of $S_F^{2k}$  as the fuzzy fibre $S_F^{2k-2}$ over $S^{2k}$;  
\be
S^{2k}_F ~\sim~ S^{2k}\otimes S_F^{2k-2}. 
\ee
(Here, $\sim$ denotes local equivalence.)  
Although $S_F^{2k-2}$ represents the internal non-abelian gauge space of the particle, 
the internal space is as large as the external space $S^{2k}$, and it can 
be regarded as  $(2k-2)$-brane in the enlarged space [see Fig.\ref{so2kflux}] that consists of  the external space $S^{2k}$ and the ``internal'' space $S^{2k-2}$ which membrane occupies. 
Since membrane is associated with the flux of non-abelian monopole, membrane can be considered as  a charged excitation induced by a penetration of the non-abelian flux in higher dimensions. 

\begin{itemize}
\item Emergence of $(4k-1)$D space-time and $J$-homomorphism     
\end{itemize}

Though we started from the $(2k+1)$D space-time where color particles  and the $SO(2k)$ non-abelian monopole live, we arrive at $(4k-1)$D space-time where $(2k-2)$-brane and $(2k-1)$ rank tensor monopole live. 
Mathematically, the Hopf-Whitehead $J$-homomorphism \cite{RavenelZee1985, WuZee1988, TzeNam1989, Bernevigetal2002} \footnote{
In general, $J$-homomorphism represents the homomorphism between the homotopy group of the orthogonal group and that of sphere: 
\begin{equation}
\pi_l(SO(M))~\rightarrow \pi_{l+M}(S^M).
\label{generalJhomomorphism}
\end{equation}
Eq.(\ref{nontrivialJhomo}) can be regarded as  special cases of (\ref{generalJhomomorphism}) for $l=2k-1$ and $M=2k$. 
When $l=1$, the homomorphism (\ref{generalJhomomorphism}) becomes the isomorphism: 
\begin{equation}
\pi_1(SO(M))=\pi_{M+1}(S^M), 
\end{equation}
which gives the 1st Hopf map, 
$\pi_{3}(S^2)=\pi_1(SO(2))\simeq \mathbb{Z}$, for $M=2$.  
The other two Hopf maps are also obtained as the  $J$-homomorphim (\ref{nontrivialJhomo}) for $k=2, 4$:  
\begin{align}
&\pi_3(SO(4))\simeq \mathbb{Z}\oplus\mathbb{Z}~\rightarrow~ \pi_7(S^4)\simeq \mathbb{Z}\oplus\mathbb{Z}_{12},\nn\\
&\pi_7(SO(8))\simeq \mathbb{Z}\oplus\mathbb{Z}~\rightarrow~ \pi_{15}(S^8)\simeq \mathbb{Z}\oplus\mathbb{Z}_{120}. \label{3rdhopfmap}
\end{align}\label{hopfmaps}
}
accounts for the intimate connection between the 
$(2k+1)$D space(-time) and the $(4k-1)$D space(-time): 
\be
\pi_{2k-1}(SO(2k))\simeq \mathbb{Z}~~\rightarrow ~~\pi_{4k-1}(S^{2k})\simeq \mathbb{Z}. 
\label{nontrivialJhomo}
\ee
The left homotopy is related to the $SO(2k)$ monopole at the origin of $(2k+1)$D space and describes the non-trivial winding from the equator of $S^{2k}$ to the $SO(2k)$ monopole gauge group, while the right homotopy describes a non-trivial winding from $(4k-1)$ space(-time) to the base-manifold $S^{2k}$ on which $(2k-2)$-brane lives. 
In particular for $k=1$, (\ref{nontrivialJhomo}) gives 
\be
\pi_{1}(SO(2)\simeq U(1))\simeq \mathbb{Z}~~\rightarrow~~\pi_3(S^2)\simeq \mathbb{Z}.
\ee
The left homotopy guarantees the non-trivial topology of Dirac monopole bundle, while  
the right homotopy represents the 1st Hopf map 
 which is  the underlying mathematics of fractional statistics of 0-brane in 3D  space(-time) \cite{WilczekZee1983}.   The world line of the 0-brane on $S^2$ corresponds to the $S^1$ fibre on $S^2$, and the non-trivial linking of  world lines of two 0-branes indicates the topological number denoted by the 1st Hopf map 
\cite{BookBottTu}. Similarly, the non-trivial homotopy $\pi_{4k-1}(S^{2k})\simeq \mathbb{Z}$ is  related to the  fractional statistics in $(4k-1)$D space(-time) \cite{
NepomechieZee1984,TzeNam1989,HorowitzSrednicki1990}. The dimension of the object obeying the fractional statistics can readily be obtained by the following dimensional counting.  
Since the dimension of the total space(-time) is  $(4k-1)$
  and  $S^{2k}$ is the basemanifold, the remaining $(4k-1)-2k=2k-1$ dimension should be the dimension of the world volume of  the object that obeys the fractional statistics.   
Indeed the dimension of  $(2k-2)$-brane world volume is $(2k-1)$ dimension, and so $(2k-2)$-branes are expected to obeys the fractional statistics.  

Another way to see $(2k-2)$-brane can obey fractional statistics is to notice the co-dimension.  
The necessary condition for the existence of fractional statistics is the co-dimension 2 where the braiding operation has non-trivial meaning.  
Indeed, the co-dimension of two (non-overlapping) $(2k-2)$-branes in $(4k-2)$ space is  2 
 [Table \ref{table:membrane}].   
 From the co-dimension, two membranes are regarded as two point particles, and the idea of fractional statistics (for particles) in 3D can similarly be applied to higher dimensions.

\begin{table}
\begin{center}
   \begin{tabular}{|c|c|c|c|c|c|c|c|c|c|c|c|c|c|}\hline
    Dim.     &  0 &  1 & 2 & $\cdots$ & $2k-2$ & $2k-1$ &$2k-2$  & $\cdots$ & $4k-4$ & $4k-3$ & $4k-2$ \\ \hline
 $M_{2k-2}$  &  $\circ$   & $\circ$   & $\circ$  & $\cdots$  & $\circ$  & & &  & &  &                          \\ \hline 
 $M_{2k-2}$ &  $\circ$    &   &  & & & $\circ$  & $\circ$  & $\cdots$    & $\circ$ &  &                   \\ \hline  
    \end{tabular}       
\end{center}
\caption{Two non-overlapping $(2k-2)$-branes in $(4k-1)$D space-time. 
From the co-dimension 2, the two $(2k-2)$-branes are regarded as two point particles.} 
\label{table:membrane}
\end{table}

\begin{itemize}
\item Physical realization of fractional statistics    
\end{itemize}

The statistical transformation is physically achieved by acquiring Aharonov-Bohm phase \cite{Wilczek1982-1,Wilczek1982-2}, where the particles acquires a statistical phase during a trip around the magnetic flux.  In the fractional quantum Hall effect, the statistical phase accounts for the fractional statistics of fractionally charged quasi-particle excitation  \cite{ArovasSW1984} and also for the statistical transformation from electron to composite boson at the odd-denominator fillings \cite{Girvin-M-1987, Read-1989, Zhang-1992}. 
The statistical transformation to composite boson is elegantly described by the Chern-Simons field theory formulation \cite{Zhang-H-K-1988,Zhang-1992}.  
In higher dimensions, there are $(2k-2)$-branes coupled to the $(2k-1)$ rank tensor $U(1)$ gauge field, and the statistical transformation is generalized in higher dimensions by adopting tensor version of Chern-Simons field theory for membranes  instead of particles.  
The mathematics of linking and phase interaction mediated by tensor gauge field 
in higher dimensions have already been formulated in Refs.\cite{NepomechieZee1984,WuZee1988,TzeNam1989,Bernevigetal2002} 
[see Appendix \ref{append:linkingmem}]. Based on the results, we discuss the statistical transformation and effective field theory for the A-class topological insulator.  
We will see that A-class topological insulator can be considered as a superfluid state of composite membranes in the same way as the fractional quantum Hall effect is regarded as a superfluid state of composite bosons.

\subsection{Tensor flux attachment}\label{sec:fuxgaugequant}

\begin{table}
\begin{center}
   \begin{tabular}{|c|c|c|c|c|c|c|c|c|c|}\hline
    Dim.    & 0  &   1 &  2 &  $\cdots$  & $p$ & $p+1$& $\cdots$ & $2p+1$ & $2p+2$   \\ \hline
  $M_p$ &  $\circ$  &  $\circ$ &   $\circ$   &  $\cdots$  &  $\circ$ & &  &   & \\ \hline  
    \end{tabular}       
\end{center}
\caption{We place a static $p$-brane in  the space-time with dimension $2p+3$.  }
\label{table:positionofpbranesfit}
\end{table}

The flux attachment is achieved by applying the singular gauge transformation \cite{Wilczek1982-1, Wilczek1982-2, Girvin-M-1987}.  We first generalize this procedure in higher dimensions. 
Suppose $p$-brane occupying the dimensions from $x^0$ to $x^p$  in $D=2p+3$ [Table \ref{table:positionofpbranesfit}]. (Here, we render $p$   non-negative integers not only  even integers.)   From the remaining $(p+2)$ dimension  
($x^{p+1}, \cdots, x^{p+2}$) $p$-brane is regarded as a point-particle.    We apply the flux attachment to such a ``point-particle'' in $(p+2)$-dimensional space. 
Technically, the gauge field associated with the flux readily be obtained by a ``dimensional reduction'' of the tensor monopole gauge field (\ref{fullysymcexplis}).  
On the equator of $S^{p+2}$ $(x_{p+3}=0)$, the tensor monopole gauge field (\ref{fullysymcexplis}) is reduced to  
\be
A_{\mu_1\mu_2\cdots\mu_{p+1}}=-\Phi_p\frac{1}{\mathcal{A}(S^{p+1})} \frac{1}{r^{p+2}}\epsilon_{\mu_1\mu_2\cdots \mu_{p+2}}x^{\mu_{p+2}}, 
\label{fluxeffectivegauge}
\ee
where $\mu_1, \mu_2, \cdots, \mu_{p+2}=p+1, p+2, \cdots, 2p+2$ and $r^2=\sum_{\mu=p+1}^{2p+2}{x^{\mu}}x^{\mu}$. 
For instance, we have 
\begin{align}
&p=0~:~A_{\mu}=-\frac{\Phi_0}{2\pi r^2} \epsilon_{\mu\nu}x_{\nu}, \nn\\ 
&p=1~:~A_{\mu\nu}=-\frac{\Phi_1}{4\pi r^3} \epsilon_{\mu\nu\rho}x_{\rho}, \nn\\ 
&p=2~:~A_{\mu\nu\rho}=-\frac{\Phi_2}{2\pi^2 r^4} \epsilon_{\mu\nu\rho\sigma}x_{\sigma}. 
\end{align}
They are regarded as  the tensor gauge field on the $(p+2)$D plane [Fig.\ref{FluxMembrane}]. 
\begin{figure}[tbph]\center
\includegraphics*[width=90mm]{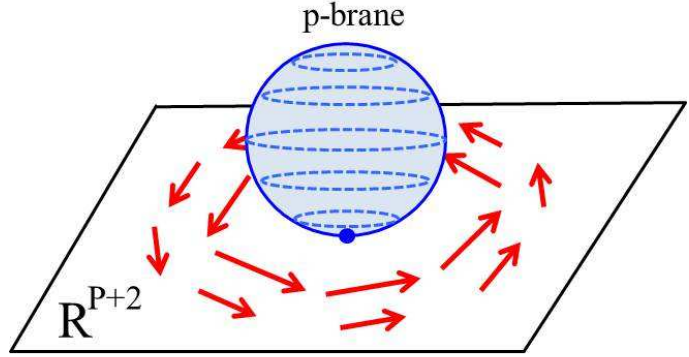}
\caption{Flux is attached to membrane and yields the tensor gauge field around the membrane.  }
\label{FluxMembrane}
\end{figure}
With use of the Green function in $(p+2)$D space\footnote{$G_{(d)}$ denotes Green function for the $d$-D Laplace equation: 
\be
\partial^2 G_{(d)}(x-y) =\delta^{d}(x-y), 
\ee
where $\partial^2=\sum_{\mu=1}^d\frac{\partial}{\partial x_{\mu}}\frac{\partial}{\partial x_{\mu}}$. Explicitly, the Green functions are given by 
\begin{align}
&d=1~~:~~G_{(1)}=\frac{1}{2}|x|~~~~~~~~~~~~~~~~~~~~~~~~~~\rightarrow~~~~~~ \partial_{x} G_{(1)}=\pm \frac{1}{2}{\text{sgn}}(x), \nn\\
&d=2~~~:~~~G_{(2)}=\frac{1}{\mathcal{A}(S^1)}\ln r ~~~~~~~~~~~~~~~~~~\rightarrow~~~~~ \partial_{\mu} G_{(2)}=\frac{1}{\mathcal{A}(S^1)}\frac{1}{r^2}x_{\mu}, \nn\\
&d \ge 3~~~:~~~G_{(d)}=-\frac{1}{(d-2) \mathcal{A}(S^{d-1})}\frac{1}{r^{d-2}}~~\rightarrow~~~~\partial_{\mu} G_{(d)}=\frac{1}{\mathcal{A}(S^{d-1})}\frac{1}{r^{d}}x_{\mu}. 
\end{align}
 },  (\ref{fluxeffectivegauge}) can be represented as 
\be
A_{\mu_1\mu_2\cdots\mu_{p+1}}=-\Phi_p\epsilon_{\mu_1\mu_2 \cdots \mu_{p+2}}\partial_{\mu_{p+2}}G_{(p+2)},  
\label{gaugefieldsingutensor}
\ee
which takes the form of ``pure gauge'': 
\be
A_{\mu_1\mu_2\cdots\mu_{p+1}}=\frac{1}{p!}\partial_{[\mu_1} \Lambda_{\mu_2 \cdots \mu_{p+1}]}, 
\ee
where $\Lambda_{\mu_1 \mu_2 \cdots \mu_{p+1}}$ is formally expressed as  
\be
\Lambda_{\mu_1 \mu_2 \cdots \mu_{p}}=(-1)^{p+1}{\Phi_p} \epsilon_{\mu_1 \mu_2 \cdots \mu_{p+2}}\partial_{\mu_{p+1}} \frac{1}{\partial^2}\partial_{\mu_{p+2}}G_{(p+2)}. 
\ee
The corresponding field strength 
\be
F_{\mu_1\mu_2\cdots \mu_{p+2}}=\frac{1}{(p+1)!}\partial_{[\mu_1}A_{\mu_2\mu_3\cdots \mu_{p+2]}}
\ee
is evaluated as 
\be
F_{\mu_1\mu_2 \cdots \mu_{p+2}}=\epsilon_{\mu_1\mu_2\cdots \mu_{p+2}}B(x), 
\ee
where $B$ represents the flux-like magnetic field:  
\be
B(x)=\Phi_p\cdot\delta^{p+2}(x). 
\ee
$\Phi_p$ stands for the strength of the flux. 
When a $p$-brane with charge $e_p$ moves around  the flux, the $p$-brane acquires the phase: 
\be
e^{i e_p \oint_{S^1 \times M_p} A}=e^{i e_p \int_{D_2 \times M_p} B}=e^{i e_p \Phi_p},  
\ee
where $M_p$ denotes the configuration of $p$-brane. 
The phase should be 1: 
\be
e^{i e_p\Phi_p}=e^{2\pi i n},  
\ee
and then $\Phi_p$ is quantized as 
\be
\Phi_p=\frac{2\pi}{e_p}n, 
\label{phiquntizeinteger}
\ee
with integer $n$.  
Hence, the minimum unit of flux is given by\footnote{ Eq.(\ref{deltaequalinverse})  is consistent with the result of the charge quantization of monopole: 
\be
e_p e_{D-p-4}=2\pi n. 
\label{membranequantiz}
\ee
This manifests Dirac quantization condition between $p$ and $(D-p-4)$-branes ($D$ is the space-time dimension).  
Since non-overlapping $p$ and $(D-p-4)$ branes occupy $D-3$ spacial dimensions,   
from the co-dimension 3,  the $p$ and $(D-p-4)$-branes are regarded as point-like objects, and so we can apply the ordinary Dirac quantization condition to  $p$ and $(D-p-4)$ branes (\ref{membranequantiz}) in the same way as electron and monopole in 3D.    
Consequently, the minimum unit of the $(D-p-4)$-brane charge is derived as    
\be
\Delta e_{D-p-4}=\frac{2\pi}{e_p}, 
\label{minidd4pdiff}
\ee
which is consistent with (\ref{phiquntizeinteger}).  
} 
\be
\hat{\Phi}_p=\frac{2\pi}{e_p}. 
\label{deltaequalinverse}
\ee
Let us consider a (composite) $p$-brane that carries $\kappa$ fluxes:  
\be
Q_p=\kappa \hat{\Phi}_p,  
\label{fluxattachglobal}
\ee
where  $Q_p$ denotes the $p$-brane charge. In the  $(D-p-1)$-dimensional space perpendicular to $p$-brane, (\ref{fluxattachglobal}) can locally be rewritten as  
\be
\rho_{\text{eff}}(x_{\perp})=\frac{1}{e_p} B_{\text{eff}}(x_{\perp}), 
\label{effectperp}
\ee
where  
\be
e_p\rho_{\text{eff}}(x_{\perp})
=Q_p~\delta^{(p+2)}(x_{\perp}),~~~~B_{\text{eff}}(x_{\perp})
=\kappa\hat{\Phi}_p~\delta^{(p+2)}(x_{\perp}), 
\ee
with $x_{\perp}^{\mu}=(x^{p+1}, x^{p+2}, \cdots, x^{D-1})$. 
Furthermore, one may readily derive (\ref{effectperp}) by integrating 
\be
\rho(x)=\frac{1}{e_p} B(x), 
\label{totallicalrelfluxatt}
\ee
over the space parallel to $p$-brane, $x_{\parallel}=(x^1, x^2, \cdots, x^p)$, with use of 
\be
e_p\rho_{\text{eff}}(x_{\perp})=e_p\int d^p x_{\parallel} ~\rho(x), ~~~~~
B_{\text{eff}}(x_{\perp})=\int  d^px_{\parallel} ~B(x). 
\ee
Here, $J_{\mu_1\mu_2\cdots \mu_{p+1}}(x)$ denotes the $p$-brane current\footnote{The explicit form of the membrane current is given as follows. 
We place $p$-brane in the dimensions, $(x^1,x^2,\cdots,x^p)$, 
 and parameterize the coordinates of membrane as 
\begin{align}
&x^{\mu}_{\parallel}=X^{\mu}(\sigma), ~~~(\mu=1,2,\cdots,p), \nn\\
&x^{\mu}_{\perp}=0, ~~~~~~~~~~(\mu=p+1,\cdots,D-1) 
\end{align}
where $\sigma=(\sigma^1, \sigma^2, \cdots, \sigma^p)$ denotes the intrinsic coordinates of the the $p$-brane. 
Non-vanishing component of $p$-brane current is given by 
\be
J^{012\cdots p}(x)=\int  d^p\sigma ~\det(\frac{\partial X}{\partial \sigma})~\delta^{(D)}(x-X(\sigma))=\delta^{(D-p-1)}(x_{\perp})\int d^p \sigma ~\det(\frac{\partial X}{\partial \sigma})~ \delta^{(p)}(x_{\parallel}-X(\sigma )), 
\ee
and the total charge $Q_p$ is evaluated as 
\be
Q_p=e_p\int d^{D-1}x~ J^{012\cdots D-1}(x)=e_p\int d^p\sigma \det(\frac{\partial X}{\partial \sigma})=e_p \cdot 
V_p.   
\ee
Here, $V_p$ denotes the volume of the $p$-brane,  
$V_p\equiv \int d^p\sigma \det(\frac{\partial X}{\partial \sigma}).$  
} 
and  $\rho(x)$ and $B(x)$ are given by 
\be
\rho(x)=J^{012\cdots p-1}(x), ~~~
B(x)=F_{p+1, p+2, \cdots, 2p+2}(x). 
\ee  
Consequently,  one can find the covariant expression for (\ref{totallicalrelfluxatt}):    
\be
J_{\mu_1\mu_2\cdots \mu_{p+1}}(x)=\frac{1}{(p+2)!} \frac{1}{e_p}\epsilon_{\mu_1\mu_2\cdots\mu_{2p+3}}F^{\mu_{p+2}\cdots \mu_{2p+3}}(x).  
\label{flxuattachcompeq}
\ee
This realizes the tensor flux attachment to $p$-brane in $(2p+3)$D space(-time), and is a natural generalization of the flux attachment in 3D space(-time): 
\be
J_{\mu}=\frac{1}{2e_0}\epsilon_{\mu\nu\rho}F^{\nu\rho}. 
\ee


\subsection{$(2k-2)$-brane as the $SO(2k+1)$ skyrmion}

In the realization of the fractional statistics of the $SO(3)$ nonlinear model in (2+1)D \cite{WilczekZee1983,BowickKarabaliWijewardhana1986},  the statistical gauge field is coupled to the $SO(3)$ skyrmion topological current. 
The underlying mathematics of the $SO(3)$ skyrmion is given by the 1st Hopf map \cite{WilczekZee1983}, where the target space $S^2$ (\ref{1sthopfexp}) corresponds to the field manifold of skyrmion.  Since both of the $SO(3)$ non-linear sigma model and the Haldane's two-sphere are based on the 1st Hopf map, 
the mathematical structure of the $SO(3)$ non-linear sigma model is quite similar to that of the Haldane's two-sphere \cite{haldane1983};  The $\it{internal}$ field manifold of the $SO(3)$ skyrmion is $S^2$  and  the ``hidden'' local symmetry  is $U(1)$, while in the  Haldane's two-sphere the $\it{external}$ space is $S^2$ and the gauge symmetry is $U(1)$.   
 Thus interestingly, we can ``interchange'' the $SO(3)$ non-linear sigma model and the Haldane's two-sphere by exchanging external and internal spaces.  
The authors in \cite{WuZee1988, NepomechieZee1984, TzeNam1989, DemlerZhang1998}  adopted the 2nd Hopf map (and the 3rd Hopf map also) to construct the $SO(5)$  non-linear sigma model for 2-brane on a four-sphere.  
We further apply this idea to construct the non-linear sigma model for  membrane of higher dimensional quantum Hall effect. 
 Since $2k$D quantum Hall effect accommodates  the ``internal''  $(2k-2)$-brane on the external space $S^{2k}$, the corresponding non-linear sigma model is the $SO(2k+1)$ non-linear sigma model realizing a skyrmion solution  spatially extended over $S^{2k-2}$ with   $S^{2k}$ internal space.  
The internal space coordinates of the $SO(2k+1)$ skyrmion are given by
\begin{equation}
n=\sum_{a=1}^{2k+1}n_a \gamma_a
\end{equation}
where $n$ is subject to the condition of $S^{2k}$:  
\begin{equation}
n^2=\sum_{a=1}^{2k+1}n_an_a=1.  
\end{equation}
Following to the Derrick's theorem, there do not exist static soliton solutions in the scalar field theory whose Lagrangian only consists of the second order kinetic term,   $\text{tr}(\partial^{\mu}n)^{\dagger}(\partial_{\mu}n)$, and  self-interaction potential in the space-time dimension larger than 2.  However, there are at least two ways to evade the Derrick'e theorem. One is to include  an extra interaction term to stabilize the soliton configuration, and the other is to adopt a higher derivative kinetic term \cite{Kafiev1981}.  
Here we just suppose that the skyrmion configuration is stabilized by taking some method to evade the theorem.

The $(2k-2)$-brane charge is given by  the $SO(2k+1)$ skyrmion topological number, 
\be
\pi_{2k}(S^{2k})\simeq \mathbb{Z}.
\ee 
In $(4k-1)$D space-time,  
 the $SO(2k+1)$ skyrmion or $(2k-2)$-brane current is constructed as 
\begin{align}
&J^{\mu_1\mu_2\cdots\mu_{2k-1}}=\frac{1}{(2k)!} \epsilon_{a_1a_2\cdots a_{2k+1}}\epsilon^{\mu_1\mu_2\cdots \mu_{4k-1}}n^{a_1} {\partial_{\mu_{2k}}}n^{a_2}  {\partial_{\mu_{2k+1}}}n^{a_3}\cdots {\partial_{\mu_{4k-1}}}n^{a_{2k+1}},  
\label{topologicalcurrentmenbrane}
\end{align}
where $\partial_{\mu}\equiv \frac{\partial}{\partial x_{\mu}}$ $(\mu=0,1,2,\cdots, 4k-2)$.   
In the differential form,  (\ref{topologicalcurrentmenbrane}) is simply represented as
\footnote{$(D-p-1)$ form current $J$ is introduced as 
\be
(*J)_{p+1}=\frac{1}{(p+1)!}J_{\mu_1 \mu_2 \cdots \mu_{p+1}}dx^{\mu_1} dx^{\mu_2}  \cdots  dx^{\mu_{p+1}}, 
\ee
and so  
\be
J_{D-p-1}
= \frac{1}{(p+1)! (D-p-1)!} \epsilon^{\mu_1 \mu_2 \cdots \mu_D}J_{\mu_1 \mu_2 \cdots \mu_{p+1}}dx_{\mu_{p+2}} dx_{\mu_{p+3}} \cdots  dx_{\mu_D}.   
\ee
}  
\be
J_{2k}=\frac{1}{(2i)^k}\text{tr}(n(dn)^{2k}), 
\label{2kformmembranecurrent}
\ee
where 
$\text{tr}(\gamma_{\mu_{2k+1}}\gamma_{\mu_1}\gamma_{\mu_2}\cdots \gamma_{\mu_{2k}})=(2i)^k\epsilon_{\mu_1\mu_2\mu_3\cdots\mu_{2k+1}} $ was used.  
The topological number of the $SO(2k+1)$ skyrmion is given by  
\be
N=\frac{1}{\mathcal{A}(S^{2k})}\int_{S^{2k}} J_{2k}. 
\ee

\subsection{Flux cancellation and tensor Chern-Simons theory}\label{subsec:fluxcantensorCS}

 Topological features of the fractional quantum Hall effect are nicely captured by the Chern-Simons effective field theory\cite{Zhang-H-K-1988,Zhang-1992}. The  Chern-Simons field  is introduced to cancel the external magnetic field, and 
 the odd number Chern-Simons fluxes attachment transmutes electron to composite boson. 
In 2D, both of the external magnetic field and the Chern-Simons field are $U(1)$, and then the relation for flux cancellation is rather trivial 
\be
\text{2D}~:~A -C_{1}=0.  
\ee
Meanwhile in higher dimensions, we have to deal with the non-abelian external field and  membranes. One may wonder how we can incorporate these two objects to generalize the flux cancellation.    
The non-abelian and tensor monopole correspondence (\ref{nonabeliantensorcorres}) gives a crucial hint: 
The non-abelian  gauge field is ``equivalent'' to the $U(1)$ tensor gauge field.  This suggests that the cancellation of the external $\it{non}$-$\it{abelian}$ gauge field by  $\it{abelian}$ gauge (tensor) field is possible.   
We thus consider the $U(1)$ Chern-Simons tensor flux attachment to membrane, and then the flux cancellation condition can be generalized in higher dimensions as       
\be
\text{$2k$D}~:~\text{tr}(L_{\text{CS}}^{(2k-1)}[A]) -C_{2k-1}=0. 
\label{2kDfluxcanllelation}
\ee
For instance, (\ref{2kDfluxcanllelation}) yields   
\begin{align}
&\text{4D}~:~\text{tr}(AdA+\frac{2}{3}iA^3)-C_{3}=0, \nn\\
&\text{6D}~:~\text{tr}(A(dA)^2+\frac{3}{2}iA^3dA-\frac{3}{5}A^5)-C_{5}=0. 
\end{align}
Since the membranes are the fundamental objects in A-class topological insulator, it is natural to reformulate the theory by using the membrane degrees of freedom. 
We propose a tensor type Chern-Simons field theory as the effective field theory for A-class topological insulator\footnote{In \cite{Bernevigetal2002}, the authors adopted the ordinary $\it{vector}$ (6+1)D $U(1)$  Chern-Simons theory as an effective field theory for 4D quantum Hall effect, which describe 0-branes rather than membranes.}:  
\be
S=e_{2k-2}\int_{4k-1} C_{2k-1}J_{2k}+\frac{\kappa}{2}\int_{4k-1}C_{2k-1}G_{2k}, \label{totalcsaction}
\ee
where $J_{2k}$ denotes the $(2k-2)$-brane current $(\ref{2kformmembranecurrent})$ and 
$G_{2k}=dC_{2k-1}.$   
The tensor Chern-Simons action  yields the tensor flux attachment (\ref{flxuattachcompeq})  and is equivalent to the one  used in  the analysis of linking of membrane currents \cite{TzeNam1989}.  
The  Chern-Simons coupling is given by 
\be
\kappa=\frac{1}{\hat{\Phi}_{2k-2}} \nu_{2k}=\frac{e_{2k-2}}{2\pi} \frac{1}{m^k},  
\ee
where $\hat{\Phi}_{2k-2}$ denotes the unit-flux (\ref{deltaequalinverse}) and $\nu_{2k}$ stands for the filling factor of $(2k-2)$-branes (\ref{fillingmembr}).  
Notice that while the original space-time dimension  is $(2k+1)$, the tensor Chern-Simons theory is defined in the enlarged $(4k-1)$D space(-time)  
[Table \ref{table:chern-simonstheo}] as 
\begin{table}
\begin{center}
   \begin{tabular}{|c|c|c|c|c|c|c|c|c|c|c|c|c|}\hline
 The original space-time D.  & ~3~ & ~5~ & ~7~ & ~9~  & 11 & 13 & 15 & 17  & 19 &  $\cdots$ \\ \hline
 The emergent space-time D.      & ~3~ & ~7~  & 11   & 15 & 19 & 23 & 27 & 31 & 35 &  $\cdots$ \\ \hline
    \end{tabular}       
\end{center}
\caption{ The effective field theory of A-class topological insulators  in 
$(2k+1)$D space-time is given by the tensor Chern-Simons theory  of  $(2k-2)$-branes in $(4k-1)$D  space-time. } 
\label{table:chern-simonstheo}
\end{table}
 consistent with  the observation in Sec.\ref{subsec:non-commutativegeo}.  
 It is also noted that the tensor Chern-Simons theory is not defined in arbitrary odd dimensional space but only  in $(4k-1)$D space.  In $(4k-3)$D space, the tensor Chern-Simons term always vanishes due to even rank Chern-Simons tensor field. 
 
Since there does not exist the kinetic term in the action, $C_{2k-1}$ is not a dynamical field but an auxiliary field determined by the equations of motion
\footnote{In component representation, (\ref{totalcsaction}) and (\ref{equationforcp+1}) are respectively expressed as 
\begin{align}
&S=\frac{1}{(2k-1)!}\int d^{4k-1}x~\biggl(-e_{2k-2} J_{\mu_1\mu_2\cdots \mu_{2k-1}}C^{\mu_1\mu_2\cdots \mu_{2k-1}}+\frac{\kappa}{2(2k)!} \epsilon^{\mu_1\mu_2\cdots \mu_{4k-1}}C_{\mu_1\mu_2\cdots \mu_{2k-1}}G_{\mu_{2k}\mu_{2k+1}\cdots \mu_{4k-1}}\biggr), 
\label{tensorchernsimonslagcoup}\\
&J^{\mu_1\mu_2\cdots\mu_{2k-1}}=-\frac{1}{(2k)!}\frac{\kappa}{e_{2k-2}}\epsilon^{\mu_1\mu_2\cdots \mu_{4k-1}}G_{\mu_{2k}\mu_{2k+1}\cdots\mu_{4k-1}}.  
\end{align}
} 
\be
J_{2k}=-\frac{\kappa}{e_{2k-2}} G_{2k}. 
\label{equationforcp+1}
\ee
In the space-time components, (\ref{equationforcp+1}) can be written as 
\begin{subequations}
\begin{align}
&J^{i_1i_2\cdots i_{2k-2} 0} =-\frac{\kappa}{e_{2k-2}}B^{i_i i_2\cdots i_{2k-2}}, \label{halleffectgene1} \\
&J^{i_1i_2\cdots i_{2k-1}} =\frac{1}{(2k)!}\frac{\kappa}{e_{2k-2}}\epsilon^{i_1i_2\cdots i_{4k-2}}E_{i_{2k}\cdots i_{4k-2}}, \label{halleffectgene2}
\end{align}
\end{subequations}
where $\epsilon^{i_1i_2\cdots i_{4k-2}}\equiv \epsilon^{i_1i_2\cdots i_{4k-2} 0} $ and 
\bse
\begin{align}
&E_{i_1 i_2 \cdots i_{2k-1}}\equiv G_{0 i_1 i_2 \cdots i_{2k-1}}=\frac{1}{(2k)!}\partial_{[0}C_{i_1 i_2 \cdots i_{2k-1}]}, \\
&B^{i_1i_2\cdots i_{2k-2}}=\frac{1}{(2k)!}\epsilon^{i_1i_2\cdots i_{4k-2}}G_{i_{2k-1}\cdots i_{4k-2}}. 
\end{align}
\ese
(\ref{halleffectgene1}) realizes the generalized flux attachment for membrane (\ref{flxuattachcompeq}) and suggests that the membrane with unit charge $e_{2k-2}$ carries $m^k$ fluxes in unit of $\hat{\Phi}_{2k-2}$. 
 Meanwhile (\ref{halleffectgene2}) gives a generalization of the Hall effect. 
From the antisymmetric property of the epsilon tensor, we have 
\be
E_{i_1 i_2 \cdots i_{2k-1}} J^{i_1 i_2 \cdots i_{2k-1}}=- E_{i_1 i_2 \cdots i_{2k-1}} J^{i_1 i_2 \cdots i_{2k-1}}=0, 
\label{hallorthogonalitypm}
\ee
which denotes a generalization of the orthogonality between Hall current and electric field.

\subsection{Composite membrane and fractional charge}\label{sec:chargemembrances}

 Integration of the Chern-Simons field in the tensor Chern-Simons action gives a generalized  Gauss-Hopf linking between two membrane world volumes, which can   
alternatively be understood as the winding number from the two higher dimensional  ``tori'' to a higher dimensional sphere [see \cite{TzeNam1989} or Appendix \ref{append:linkingmem}]: 
\be 
 (S^{2k-2}\times S^1) \times   (S^{2k-2}\times S^1) ~\rightarrow S^{4k-1}.  
 \label{windinghighspheres}
\ee 
From (\ref{windinghighspheres}), it is obvious that the non-trivial winding exists for arbitrary $k$, and so does   the linking. 
Even though the membrane statistics is related to the linking, it does not necessarily mean that  membranes obey the fractional statistics.    
For instance in quantum Hall effect, for quasi-excitation to be anyonic, the fractional charge is essential \cite{ArovasSW1984}.  
Similarly, for statistical transmutation from electron to (composite) boson, the odd number flux attachment is crucial. 

First, we consider the composite boson counterpart in  A-class topological insulators. At $\nu=1/m^k$, $m^k$ fluxes are attached to the membrane and the membrane becomes a 
composite object of the original membrane and the fluxes. The original statistics of the membrane is 
fermionic since at $\nu=1$ membrane corresponds to ``quarks'' with color degrees of freedom.     
The statistics of the composite membrane is derived by evaluating the phase interaction between two composite membranes.  
Under the interchange, the composite membranes acquires the following statistical phase 
\be
e^{i\frac{1}{2}e_{2k-2} \oint A  }  =e^{i\pi m^k  }=-1, 
\ee
where we used $ \oint A  =m^k\Phi_{2k-2}$ ($\Phi_{2k-2}=\frac{2\pi}{e_{2k-2}}$)  and $m$ is odd so is  $m^k$. Since the composite membrane acquires the extra minus sign under the interchange of two composite membranes, the flux attachment induces the statistical transformation of membrane from fermion to boson, and the composite membrane obeys the Bose statistics.  
Notice that such transmutation is only possible for the special filling fraction when the inverse of the filling fraction is odd (${m^k}$).   In the same way as the fractional quantum Hall effect at $\nu=1/m$ is regarded as a condensation of composite bosons, the A-class topological insulator at $\nu_{2k}=1/m^k$ may be considered as a superfluid state of composite membranes.  
Next let us discuss the statistics of membrane excitation. We first need to specify the membrane charge. 
 When the monopole charge is $I/2$,  the number of states on $S^{2k}$ is given by 
\be
\sim I^k,  
\ee
and for the filling $\nu=1$ the $(2k-2)$-brane with unit charge 
 $e_{2k-2}$, 
 occupies each state.  
When the monopole charge change as  $I'=mI$, the number of states becomes to  
\be
{I'}^{k}=m^k I^k. 
\ee
In other words, each state occupied by membrane is ``split'' to $m^k$ states, and 
 so does the membrane charge. Hence at $\nu_{2k}={1}/{m^k}$,  
the fractional charge of $(2k-2)$-brane is given by\footnote{
Eq.(\ref{fractionalcharge2k-2}) can also be derived from the perspective of $0$-branes. 
When the monopole charge is $I/2$, the $(2k-2)$-brane is made of  
$I^{\frac{1}{2}k(k-1)}$
 0-branes, and then $(2k-2)$-brane charge is expressed by 
\be
e_{2k-2}=\kappa(k)\cdot I^{\frac{1}{2}k(k-1)}e_0,  
\ee
where $\kappa(k)$ is a coefficient of dimension of ${(\text{mass})^{2k-2}}$.  
At $I'=mI$, the 0-brane charge becomes to 
\be
e_0'=\frac{1}{m^{\frac{1}{2}k(k+1)}}e_0, 
\ee
and so the $(2k-2)$-brane charge is derived as   
\be
e'_{2k-2}=\kappa(k)\cdot {I'}^{\frac{1}{2}k(k-1)}e_0'=\kappa(k)\cdot \frac{1}{m^{\frac{1}{2}k(k+1)}}{I'}^{\frac{1}{2}k(k-1)}e_0=\frac{1}{m^k}e_{2k-2}. 
\ee
}: 
\be
e'_{2k-2}=\frac{1}{{I'}^k}I^{k}e_{2k-2}=\frac{1}{m^k}e_{2k-2}. 
\label{fractionalcharge2k-2}
\ee
Since  the $(2k-2)$-brane excitation is induced by the flux penetration, $(2k-2)$-brane excitation is a ``composite'' of the fractional charge $e'_{2k-2}$ and the unit flux $\hat{\Phi}_p={2\pi}/{e_{2k-2}}$. Therefore, the geometrical phase which  a fractionally charged $(2k-2)$-brane acquires during the round trip around another $(2k-2)$-brane  is given by   
\be
e^{ie'_{2k-2} \oint A  } =e^{ie'_{2k-2}\hat{\Phi}_p}=e^{  2\pi i\frac{e'_{2k-2}}{e_{2k-2}}}=e^{ \frac{2\pi}{m^k}i}.
\ee
Thus, the statistical phase of membrane excitation is $2\pi\nu_{2k}$, and hence membrane excitations are anyonic.

\subsection{Dimensional hierarchy and analogies to string theory}\label{sec:dimensionalhier}

Analogies between the A-class topological insulator and the string theory will be transparent in analyses  of membrane properties.    
 According to the Haldane-Halperin picture \cite{haldane1983,Halperin1984}, quasi-particles condense on the parent quantum Hall liquid to generate a new incompressible liquid and the filling factor exhibits a hierarchical structure called Haldane-Halperin hierarchy.   
Similarly in A-class topological insulator, membrane excitations are expected to condense to form a new incompressible liquid,  and the filling factor will exhibit a generalized  Haldane-Halperin like hierarchy: 
\be
\nu_{2k}=\frac{1}{m^k\pm \frac{1}{(2p_1)^k\pm\frac{1}{(2p_2)^k\pm\cdots}}}, 
\label{haldahalpmemb}
\ee
where each of $p_1, p_2, \cdots$ denotes a natural number.  
Apart from the Halperin-Haldane hierarchy, the membranes exhibit a unique type of condensation -- the dimensional hierarchy \cite{HasebeKimura2003, Hasebe2010}, which reflects the special dimensional pattern of A-class topological insulator.    
From (\ref{iterativedegeneracylll}), one may find that there is a relation between  $2k$ and $(2k-2)$D lowest Landau level degeneracies:   
\be
D_{LLL}(k,I)
~\sim~ I^{k}D_{LLL}(k-1,I), 
\label{iterativedegeneracy}
\ee
and then 
\be
D_{LLL}(k,I) ~~~\sim 
~~~I^k \cdot I^{k-1} \cdot I^{k-2} \cdots I^2 \cdot I
=I^{\frac{1}{2}k(k+1)}.
\label{degeneLLL}
\ee
Eq.(\ref{degeneLLL}) implies a hierarchy ranging over dimensions.  This feature can intuitively be understood by  the following simple explanations.  
Each of the $SO(2k)$ monopole fluxes on $S^{2k}$ occupies an area $\ell_B^{2k}=(\alpha r)^k=
{(2r^2/I)}^{k}$, 
and the number of fluxes on $S^{2k}$ is given by  $\sim r^{2k}/\ell_B^{2k}\sim I^k $. 
Since the $SO(2k)$ non-abelian flux is equivalent to $(2k-2)$-brane, one may say $(2k-2)$-brane  occupies the same area $\ell_B^{2k}$ and $\sim I^{k}$ is the number of $(2k-2)$-branes.   
Similarly, on $S^{2k-2}$, there are $(2k-4)$-branes each of which  occupies the area $l_B^{2k-2}$, and the total number of $(2k-4)$-branes is  $\sim I^{k-1}$.    By repeating this  iteration from $2k$D to the lowest dimension $2$D, we obtain the formula (\ref{degeneLLL}). The corresponding filling factor (for 0-brane) is given by 
\be
\nu=\frac{1}{m}\frac{1}{m^2}\frac{1}{m^3}\cdots \frac{1}{m^{k-1}}\frac{1}{m^k}=\frac{1}{m^{\frac{1}{2}m(m+1)}}. 
\label{simpestdimhier}
\ee
Similar to the Haldane-Halperin hierarchy, such a hierarchical structure may imply a particular condensation property of  membranes.  
 One may see the formula from low dimension to say  low dimensional membranes gather to  form  a higher dimensional incompressible liquid of membranes 
[Fig.\ref{DimHierarchy}]. This is the physical interpretation of the dimensional hierarchy of the filling fraction (\ref{simpestdimhier}). 
Most general total filling factor will be given by the combination of (\ref{haldahalpmemb})  and  (\ref{simpestdimhier}): 
\be
\nu=\nu_2 \nu_4 \cdots \nu_{2k}=
\frac{1}{m\pm \frac{1}{2p_1\pm\frac{1}{2p_2\pm\cdots}}}
\cdot 
\frac{1}{m^2\pm \frac{1}{(2p_1)^2\pm\frac{1}{(2p_2)^2\pm\cdots}}}
\cdots
\frac{1}{m^k\pm \frac{1}{(2p_1)^k\pm\frac{1}{(2p_2)^k\pm\cdots}}}. 
\label{totalfilling0branes}
\ee

\begin{figure}[tbph]\center
\hspace{8cm}\includegraphics*[width=145mm]{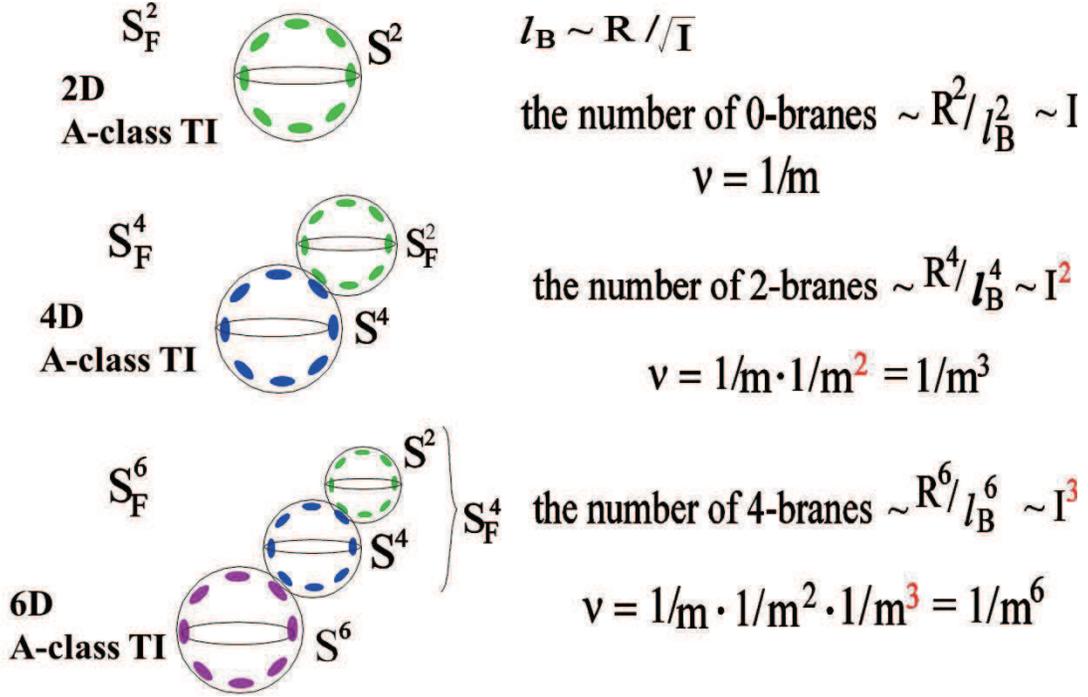}
\caption{Low dimensional membranes condense    
 to form a  higher dimensional membrane. Since the membrane itself describes fuzzy sphere or A-class topological insulator, one may alternatively interpret this phenomena as  
the dimensional hierarchy of A-class topological insulator.   }
\label{DimHierarchy}
\end{figure}

Since $\nu_{2}, \nu_{4}, \cdots, \nu_{2k}$ are equally treated  in (\ref{totalfilling0branes}), one can arbitrarily interchange $\nu$s. The interchangeability of the filling fractions in different dimensions suggests a ``democratic'' property of A-class topological insulator, $i.e.$  
equivalence between membranes of different dimensions.    
This may immediately remind the brane democracy of string theory; any D-brane can be a starting point to construct another D-brane in different dimensions \cite{Townsend-1995}.   
Thus, the dimensional hierarchy --  membranes condense to make an incompressible liquid -- is  regarded as a physical realization of the brane democracy. 
The index theorem also suggests  close relations between the A-class topological insular and the string theory. The index theorem tells that  the lowest Landau level degeneracy, $D_{LLL}(k-1, I)$, is equal to  the $(k-1)$th Chern-number,  
$c_{k-1}(I+1)$.   
This equality means that the $(k-1)$th Chern number  is identical to the  $(2k-2)$-brane charge, since the number of 0-branes is given by the lowest Landau level degeneracy.  
Analogous phenomena have  been reported in the context of Myers effect of string theory \cite{Myers-1999} where  low dimensional D-branes on higher dimensional D-brane are regarded as magnetic fluxes of monopole.    
In particular, Kimura found that the number of D0-branes that constitute a spherical  D$(2k-2)$-brane  is given by the $(k-1)$th Chern-number of non-abelian monopole \cite{Kimura2004}. 
The fact that the membrane charge is equal to the lowest Landau level degeneracy $i.e.$ the number of the fundamental elements, implies that  membranes themselves should be identified with the fundamental elements of the  space(-time). 
 This observation again reminds the idea of the matrix theory \cite{Banks-etal-1997,Ishibashi-etal-1997} in which  the D0 (D$-1$)) branes constitute the space(-time) and the spacial coordinates are represented by matrices.   
It is quite interesting that the ideas of the string theory can be understood in the context of  topological insulators.

\section{Summary and Discussions}\label{sec:summary}

We discussed physical realization of the quantum Nambu geometry in the context of A-class topological insulator. 
As the higher dimensional fuzzy sphere has two different formulations,  A-class topological insulator has two physically different realizations, one of which is the non-abelian monopole realization and the other is the tensor monopole realization. 
We established the connection between these two kinds of monopole through the  Chern-Simons term. 
Based on the non-abelian and tensor connection, we generalized the flux attachment procedure in A-class topological insulator to construct the Chern-Simons tensor effective field theory. 
We also showed the exotic concepts in 2D quantum Hall effect can  naturally be generalized to  A-class topological insulators.   

 For convenience of readers,  we summarize the main achievements of the present work.   
 In arbitrary even dimension we established 
\begin{itemize}
\item Equality between monopole charge and the lowest Landau level degeneracy via the index theorem [Sec.\ref{sec:so2k+1landaumodel}, \ref{sec:indextheorem}] 
\item Connection between the non-abelian and tensor monopoles  [Sec.\ref{subsec:relationsfieldmoono}]
\end{itemize}
Based on the above observations, we derived 
\begin{itemize}
\item Explicit form of the tensor monopole gauge fields from the non-abelian monopole gauge fields [Sec.\ref{subsec:relationfields}]
\item Non-commutative coordinates of quantum Nambu geometry via angular momentum construction [Sec.\ref{sec:tenformononambuncg}]
\end{itemize}
Subsequently, we discussed their physical consequences in the context of A-class topological insulators: 
\begin{itemize}
\item Tensor flux attachment to membrane and its statistical phase [Sec.\ref{sec:fuxgaugequant}]
\item Higher D generalization of flux cancellation and Chern-Simons tensor field theory [Sec.\ref{subsec:fluxcantensorCS}]
\item  Fractional charge and anyonic statistics for membrane [Sec.\ref{sec:chargemembrances}]
\end{itemize}
While the original space-time of A-class topological insulators is the space-time dimension $(2k+1)$,   
the effective Chern-Simons tensor field theory lives in the enlarged $(4k-1)$ dimensional space-time. 
The edge theory and accompanied Callan Harvey mechanism based on the Chern-Simons tensor field theory may also be interesting.  

 The quantum Nambu bracket has attracted a lot of attentions in recent years since it is expected to provide an appropriate  description for  M-brane boundstate \cite{BasuHarvey2005} and plays a vital role in Bagger-Lambert-Gustavsson theory of multiple M-branes \cite{BaggerLambert20078,Gustavsson2009,Baggeretal2013}. 
Non-associative geometry associated with the quantum Nambu bracket has also been vigorously studied \cite{Mylonas-Sch-S-2013,Bakas-Lust-2013}. 
  As a pioneer of  higher dimensional quantum Hall effect and topological insulator, Zhang noted that the study of condensed matter physics may provide an alternative path to understand  exotic ideas in mathematical and particle physics \cite{Zhang-2002}.
We thus enforced his observation by demonstrating  quantum Nambu geometry in A-class topological insulators inspired by the recent works \cite{EstienneRB2012,NeupertSRChMRB2012}.  
We hope the present work  will further deepen the understanding of  non-commutative geometry and string theory as well as topological insulators.

\section*{Acknowledgments}

The author is grateful to Shoucheng Zhang for his interests in this work and accepting the author as a visiting scholar in Stanford University.  
Discussions and email correspondences with Koji Hashimoto, Shoichi Kawamoto, Taro Kimura,  Yusuke Kimura,  Mohammad M. Sheikh-Jabbari, Richard J. Szabo  and Tomohisa Takimi were helpful to develop this work. I appreciate the valuable communications with them.    
Finally, I would like to thank  Emma Akiyama and Takahide Akiyama for their warm encouragements and supports.  
Presentations and discussions during the workshops 
 ``Exotic Space-Time Geometry and Its Application'' held at Riken in Wako, Saitama, Japan, Feb. 23, 2013 and 
 ``Noncommutative Field Theory and Gravity'' held in Corfu, Greece, Sept. 8-15, 2013, were useful to complete the present  work.      
This work was partially supported by Grant-in-Aid for Young Scientists (B) (Grant No.23740212), Overseas   Dispatching Program 2013 of National College of Technology, and The Emma Project for Art and Culture.

\appendix 

\section{Fully symmetric representations of $SO(2k+1)$ and $SO(2k)$}\label{append:so2k+1gammagenefulsym}

In the $SO(2k+1)$ fully symmetric representation $\overbrace{[\frac{I}{2},\frac{I}{2},\cdots,\frac{
I}{2}]}^{k}$, the gamma matrices satisfy 
\be
\sum_{a=1}^{2k+1}G_a G_a=I(I+2k)
\ee
and 
\be
[G_{a_1}, G_{a_2}, \cdots, G_{a_{2k}}]=i^k C'(k, I)\cdot \epsilon_{a_1 a_2 \cdots a_{2k+1}} G_{a_{2k+1}}, 
\ee
where $C'(k, L)$ is given by 
\be
C'(k, I)\equiv \frac{(2k)!!(I+2k-2)!!}{I!!}.
\ee
The $SO(2k+1)$ generators are constructed as  
\be
G_{ab}=-i\frac{1}{4}[G_a,G_b]. 
\ee
$G_a$ and $G_{ab}$ satisfy  
\begin{align}
&[G_a,G_b]=4i G_{ab},\nn\\
&[G_a,G_{bc}]=  -i(\delta_{ab}G_c-\delta_{ac}G_b) \nn\\
&[G_{ab},G_{cd}]=i(\delta_{ac}G_{bd}-\delta_{ad}G_{bc}+\delta_{bd}G_{ac}-\delta_{bc}G_{ad}), 
\end{align}
which is identical to the $SO(2k+2)$ algebra.  
$X_a$ and $X_{ab}$ operators of $S_{F}^{2k}$ are constructed as 
\begin{align}
&X_a=\frac{\alpha}{2}G_a, \nn\\
&X_{ab}=\alpha G_{ab},  
\end{align}
with $\alpha=2r/I$  (\ref{defofalpha}). 
For $I=1$, $G_a$ and $G_{ab}$ are reduced to the fundamental representation, $\Gamma_a$ (\ref{gammamtrso2k+1}) and  $\Sigma_{ab}=-i\frac{1}{4}[\Gamma_a, \Gamma_b]$.    

The $SO(2k)$ group has two Weyl representations, $\Sigma^{+}_{\mu\nu}$ and $\Sigma^{-}_{\mu\nu}$ ($\mu,\nu=1,2,\cdots, 2k$).  
For the fundamental representation $I=1$, the $SO(2k)$ Weyl generators satisfy   
\begin{subequations}
\begin{align}
&\epsilon_{\mu_1\mu_2\mu_3\mu_4\cdots\mu_{2k}}\Sigma^{\pm}_{\mu_3\mu_4}\cdots\Sigma^{\pm}_{\mu_{2k-1}\mu_{2k}}=\pm\frac{(2k-2)!}{2^{k-2}}\Sigma_{\mu_1\mu_2}^{\pm},\\
&\tr(\Sigma_{\mu_1\mu_2}^{\pm}\Sigma_{\mu_2\mu_3}^{\pm})=-2^{k-3}(2k-1)\delta_{\mu_1\mu_3}.
\end{align}\label{formulageneso2kL1I=1}
\end{subequations}
and for the fully symmetric representation $\overbrace{[\frac{I}{2},\frac{I}{2},\cdots,\frac{
I}{2}]}^{k}$, 
\begin{subequations}
\begin{align}
&\epsilon_{\mu_1\mu_2\mu_3\mu_4\cdots\mu_{2k}}\Sigma^{\pm}_{\mu_3\mu_4}\cdots\Sigma_{\mu_{2k-1}\mu_{2k}}^{\pm}=\pm\frac{1}{2^{k-2}}C'(k-1, I)~\Sigma_{\mu_1\mu_2}^{\pm},\\
&\tr(\Sigma_{\mu_1\mu_2}^{\pm}\Sigma_{\mu_2\mu_3}^{\pm})=-\frac{1}{4}D_{LLL}(k-1,I)~I(2k+I-2)~\delta_{\mu_1\mu_3}.  
\end{align}\label{relationgeneralIso2k+1gene}
\end{subequations}
Here, $D_{LLL}(k-1,I)$ denotes the dimension of the $SO(2k)$ fully symmetric representation    that is equal to the dimension of the $SO(2k-1)$ fully symmetric representation (\ref{iterativedegeneracy}).    
For the fundamental representation, $G_{\mu\nu}$ and $\Sigma_{\mu\nu}^{\pm}$ are related by (\ref{sigmapmdec}), and for generic fully symmetric representation $G_{\mu\nu}$ can be represented by a block diagonal form and $\Sigma_{\mu\nu}^{\pm}$ appear in the left-up and right-down blocks:  
\be
G_{\mu\nu}=
\begin{pmatrix}
\Sigma_{\mu\nu}^+ & 0  & 0 \\ 
0 & \ddots  &  0 \\
0 & 0   & \Sigma_{\mu\nu}^-
\end{pmatrix}. 
\ee

\section{Properties of quantum Nambu bracket}\label{append:nambu-heisenberg}

For $d=n+1$ dimensional space,  (\ref{defofnambubracket}) can be written as 
\be
[X_{a_1},X_{a_2},\cdots,X_{a_n}]= \epsilon_{{a_1} {a_2}\cdots {a_n} {a_{n+1}}}\epsilon_{b_{a_1} b_{a_2}\cdots b_{a_n} {a_{n+1}}}X_{b_{a_1}}X_{b_{a_2}}\cdots X_{b_{a_{n}}}, 
\label{commuratorspecicaln}
\ee
where $a_1,a_2,a_{n+1}, b_1, b_2, b_{n+1}=1,2,\cdots,n+1$. 
For instance,   
\be
[X_{1},X_{2},\cdots,X_{n}]= \epsilon_{\mu_1\mu_2\cdots\mu_{n}}X_{\mu_1}X_{\mu_2}\cdots X_{\mu_{n}}, 
\ee
where $\mu_1,\mu_2,\cdots,\mu_{n}=1,2,\cdots,n$.  Due to the formula  
\be
\epsilon_{a_1a_2\cdots a_n a_{n+1}}\epsilon_{b_{a_1}b_{a_2}\cdots b_{a_n}a_{n+1}}= 
\text{det}
\begin{pmatrix}
\delta_{a_1 b_{a_1}} & \delta_{a_1 b_{a_2}} & \cdots & \delta_{a_1 b_{a_n}} \\
\delta_{a_2 b_{a_1}} & \delta_{a_2 b_{a_2}} & \cdots & \delta_{a_2 b_{a_n}} \\
\vdots & \vdots & \ddots & \vdots & \\
\delta_{a_n b_{a_1}} & \delta_{a_n b_{a_2}} & \cdots & \delta_{a_n b_{a_n}}
\end{pmatrix} \equiv \text{det}(\delta_{a_i b_{a_j}}) ~~~(i,j=1,2,\cdots,n), 
\ee
(\ref{commuratorspecicaln}) can be rewritten as 
\be
[X_{a_1},X_{a_2},\cdots,X_{a_n}]= \text{det}(\delta_{a_i b_{a_j}})X_{b_{a_1}}X_{b_{a_2}}\cdots X_{b_{a_{n}}}. 
\ee
It is obvious that (\ref{commuratorspecicaln}) can be represented as the commutator or the anti-commutator of the ``sub''-brackets: 
\begin{align}
&[X_{a_1},X_{a_2},\cdots,X_{a_n}]\nn\\
&= \frac{1}{m!(n-m)!}\epsilon_{a_1a_2\cdots a_{n}a_{n+1}}\epsilon_{b_{a_1}b_{a_2}\cdots b_{a_n}a_{n+1}}[ X_{b_{a_1}}, X_{b_{a_2}}, \cdots,  X_{b_{a_m}}][ X_{b_{a_{m+1}}}, \cdots ,  X_{b_{a_{n}}}]\nn\\
&= \frac{1}{2m!(n-m)!}\epsilon_{a_1a_2\cdots a_{n}a_{n+1}}\epsilon_{b_{a_1}b_{a_2}\cdots b_{a_n}a_{n+1}}[[X_{b_{a_1}}, X_{b_{a_2}}, \cdots,  X_{b_{a_m}}], [X_{b_{a_{m+1}}}, \cdots,  X_{b_{a_{n}}}]]_{(-1)^{m(n-m)}}, \nn\\
&~~~~(m\le n) 
\label{smallerbracketsdecomp}
\end{align}
where $[~~~]_+\equiv \{~~~\}$ and $[~~~]_- \equiv [~~~]$. 
Thus, the $n$ bracket has a hierarchical structure;  $n$ bracket can be decomposed to the algebra of sub-brackets.  In particular, for $n=2k$,  $2k$ bracket can be represented by  2 brackets: 
\begin{align}
&[X_{a_1},X_{a_2},\cdots,X_{a_{2k}}]= \frac{1}{2^k}\epsilon_{a_1a_2\cdots a_{2k}a_{2k+1}}\epsilon_{b_{a_1}b_{a_2}\cdots b_{a_{2k}}a_{2k+1}}[X_{b_{a_1}}, X_{b_{a_2}}][X_{b_3},X_{b_4}]\cdots [X_{b_{a_{2k-1}}}, X_{b_{a_{2k}}}]\nn\\
&= \frac{1}{2^{2k-1}}\epsilon_{a_1a_2\cdots a_{2k}a_{2k+1}}\epsilon_{b_{a_1}b_{a_2}\cdots b_{a_{2k}}a_{2k+1}}\{\{\cdots \{\{[X_{b_{a_1}}, X_{b_{a_2}}],[X_{b_{a_3}},X_{b_{a_4}}]\},[X_{b_{a_5}},X_{b_{a_6}}]\}\cdots \},[X_{b_{a_{2k-1}}}, X_{b_{a_{2k}}}]\}
\end{align}
In particular, 
\begin{align}
[X_1,X_2,\cdots,X_{2k}]&=\frac{1}{2^k}\epsilon_{\mu_1\mu_2\cdots\mu_{2k}}[X_{\mu_1},X_{\mu_2}][X_{\mu_3},X_{\mu_4}]\cdots[X_{\mu_{2k-1}},X_{\mu_{2k}}]\nn\\
&= \frac{1}{2^{2k-1}}\epsilon_{\mu_1\mu_2\cdots\mu_{2k}}\{\{\cdots\{[X_{\mu_1},X_{\mu_2}],[X_{\mu_3},X_{\mu_4}]\},\cdots\}, [X_{\mu_{2k-1}},X_{\mu_{2k}}]\}, 
\end{align}
with $\mu_1,\mu_2,\cdots,\mu_{2k}=1,2,\cdots, 2k$. 
For $k=2, 3$, we have  
\bse
\begin{align}
[X_{1},X_{2},X_{3},X_{4}]&=\frac{1}{8}\epsilon_{\mu_1 \mu_2 \mu_3 \mu_4 } \{ [X_{\mu_1}, X_{\mu_2}], [X_{\mu_3}, X_{\mu_4}]\}\nn\\
&= \{[X_1,X_2],[X_3,X_4]\}-\{[X_1,X_3],[X_2,X_4]\}+\{[X_1,X_4],[X_2,X_3]\}, \\ 
[X_{1},X_{2},X_{3},X_{4},X_{5},X_{6}]& 
=\frac{1}{96}\epsilon_{\mu_1 \mu_2 \mu_3 \mu_4\mu_5\mu_6}\{[X_{\mu_1},X_{\mu_2},X_{\mu_3},X_{\mu_4}],    [X_{\mu_5},X_{\mu_6}]  \}
\nn\\
&=\frac{1}{32}\epsilon_{\mu_1 \mu_2 \mu_3 \mu_4 \mu_{5}, \mu_6}\{ \{ [X_{\mu_1}, X_{\mu_2}], [X_{\mu_3}, X_{\mu_4}]\}, [X_{\mu_5},X_{\mu_6}]\}. 
\end{align}
\ese
In general, 
\begin{align}
&[X_{1},X_{2}\cdots,X_{{2k}}]=\frac{1}{2^2 (2k-2)!}\epsilon_{\mu_1\cdots\mu_{2k}}
\{[X_{\mu_1},X_{\mu_2},\cdots,X_{\mu_{2k-2}}],[X_{\mu_{2k-1}},X_{\mu_{2k}}]\}\nn\\
&~~~~~~=\frac{1}{2^4 (2k-4)!}\epsilon_{\mu_1\cdots\mu_{2k}}
\{\{[X_{\mu_1},X_{\mu_2},\cdots,X_{\mu_{2k-4}}],[X_{2k-3},X_{2k-2}]\}, [X_{\mu_{2k-1}},X_{\mu_{2k}}]\}\nn\\
&~~~~~~=\frac{1}{2^6 (2k-6)!}\epsilon_{\mu_1\cdots\mu_{2k}}
\{\{\{[X_{\mu_1},X_{\mu_2},\cdots,X_{\mu_{2k-6}}],[X_{\mu_{2k-5}},X_{\mu_{2k-4}}]\},[X_{\mu_{2k-3}},X_{\mu_{2k-2}}]\}, [X_{\mu_{2k-1}},X_{\mu_{2k}}]\}\nn\\
&~~~~~~=\cdots\nn\\
&~~~~~~=\frac{1}{2^{2k-1}}\epsilon_{\mu_1\cdots\mu_{2k}}
\{\cdots\{[X_{\mu_1},X_{\mu_2}],[X_{\mu_3},X_{\mu_{4}}]\},\cdots, \}, [X_{\mu_{2k-5}},X_{\mu_{2k-4}}]\},[X_{\mu_{2k-3}},X_{\mu_{2k-2}}]\}, [X_{\mu_{2k-1}},X_{\mu_{2k}}]\}. 
\label{doclong2k}
\end{align}
In covariant form,  (\ref{doclong2k}) can be expressed as   
\begin{align}
&[X_{a_1},X_{a_2}\cdots,X_{a_{2k}}]=\frac{1}{2^2 (2k-2)!}\epsilon_{a_1a_2\cdots a_{2k+1}}\epsilon_{b_{a_1}\cdots b_{a_{2k}} a_{2k+1}}
\{[X_{b_{a_1}},X_{b_{a_2}},\cdots,X_{b_{a_{2k-2}}}],[X_{b_{a_{2k-1}}},X_{b_{a_{2k}}}]\}\nn\\
&~~~~~~=\frac{1}{2^4 (2k-4)!}\epsilon_{a_1a_2\cdots a_{2k+1}}\epsilon_{b_{a_1}\cdots b_{a_{2k}} a_{2k+1}} 
\{\{[X_{b_{a_1}},X_{b_{a_2}},\cdots,X_{b_{a_{2k-4}}}],[X_{b_{a_{2k-3}}},X_{b_{a_{2k-2}}}]\}, [X_{b_{a_{2k-1}}},X_{a_{b_{2k}}}]\}\nn\\
&~~~~~~=\frac{1}{2^6 (2k-6)!}\epsilon_{a_1a_2\cdots a_{2k+1}}\epsilon_{b_{a_1}\cdots b_{a_{2k}} a_{2k+1}} \nn\\
&~~~~~~~~~~~~~\times 
\{\{\{[X_{b_{a_1}},X_{b_{a_2}},\cdots,X_{b_{a_{2k-6}}}],[X_{b_{a_{2k-5}}},X_{b_{a_{2k-4}}}]\},[X_{b_{a_{2k-3}}},X_{b_{a_{2k-2}}}]\}, [X_{b_{a_{2k-1}}},X_{b_{a_{2k}}}]\}\nn\\
&~~~~~~=\cdots\nn\\
&~~~~~~=\frac{1}{2^{2k-1}}\epsilon_{a_1a_2\cdots a_{2k+1}}\epsilon_{b_{a_1}\cdots b_{a_{2k}} a_{2k+1}} \nn\\
&~~~~~~~~~~~~~\times 
\{\cdots\{[X_{b_{a_1}},X_{b_{a_2}}],[X_{b_{a_3}},X_{b_{a_{4}}}]\},\cdots, \}, [X_{b_{a_{2k-5}}},X_{b_{a_{2k-4}}}]\},
[X_{b_{a_{2k-3}}},X_{b_{a_{2k-2}}}]\}, [X_{b_{a_{2k-1}}},X_{b_{a_{2k}}}]\}. 
\end{align}
One may find that there exists a dimensional hierarchy: 
\be
2k ~\rightarrow~ 2k-2 ~\rightarrow ~ 2k-4 ~ \rightarrow ~ \cdots ~ 4~  \rightarrow ~ 2,  
\ee
and the non-commutativity of $2k$-bracket is boiled down to its ``constituent'' algebra. Typically, when  
\be
[X_1,X_2]=[X_3,X_4]=\cdots=[X_{2k-1},X_{2k}]=i\ell^2,
\ee
the quantum Nambu geometry becomes a simple product of the two brackets:  
\be
[X_1,X_2, \cdots, X_{2k-1}, X_{2k}]=(i\ell^2)^k=i^k \ell^{2k}.
\ee

\section{Winding number for $S^{2k-1} \rightarrow SO(2k)$ and  tensor monopole charge }\label{appen:nonandu1charges} 

The non-trivial bundle topology of the $SO(2k)$ non-abelian monopole on $S^{2k}$ is represented by the homotopy: 
\begin{equation}
\pi_{2k-1}(SO(2k))\simeq \mathbb{Z}. 
\end{equation}
The corresponding Chern number is given by 
\be
c_k=\frac{1}{\mathcal{N}}\int_{S^{2k-1}}\text{tr}(-ig^{\dagger}dg)^{2k-1}, 
\ee
where $g$ denotes the transition function  on $S^{2k-1}$ which takes its value in  an $SO(2k)$ group element and  $\mathcal{N}$ is a normalization constant defined so as to give $c_k=1$ for the  isomorphic map from $S^{2k-1}$ to $SO(2k)$. 
The isomorphic map is given by  
\begin{equation}
g=x_{2k}+i\sum_{i=1}^{2k-1}\gamma_{i}x_{i}, 
\label{mapisoptrans}
\end{equation}
where   $(x_{i},x_{2k} )\in S^{2k-1}$ are subject to $\sum_{i=1}^{2k-1}x_{i}x_{i}+x_{2k}x_{2k}=1$ and $\gamma_i$ $(i=1,2,\cdots, 2k-1)$  are the  $SO(2k-1)$ gamma matrices. 
Obviously, $g^{\dagger}g=1$. Around the north-pole $x_{2k}\simeq 1$ and $x_i\simeq 0$ $(i=1,2,\cdots,2k-1)$, the transition function behaves as 
\begin{equation}
g^{\dagger}\simeq 1, ~~~dg\simeq i\sum_{\mu=1}^{2k-1}\gamma_{i}dx_{i}, 
\end{equation}
and the normalization constant $\mathcal{N}$ is evaluated as 
\begin{align}
\mathcal{N}&=\int_{S^{2k-1}}\text{tr}(-ig^{\dagger}dg)^{2k-1}\nn\\
&~\sim~ \int_{S^{2k-1}}\text{tr}(\gamma_{i}dx_{i})^{2k-1}=\int_{S^{2k-1}}dx_{i_1}dx_{i_2}\cdots dx_{i_{2k-1}}\text{tr}(\gamma_{i_1}\gamma_{i_2}\cdots \gamma_{i_{2k-1}})\nn\\
&=(i)^{k-1}2^{k-1}(2k-1)! \mathcal{A}(S^{2k-1}), 
\end{align}
where we used 
\begin{align}
&\gamma_{i_1}\gamma_{i_2}\cdots \gamma_{i_{2k-1}}=(i)^{k-1}\epsilon_{i_1i_2\cdots i_{2k-1}}\bold{1}_{2^{k-1}}\nn\\
&dx_{i_1}dx_{i_2}\cdots dx_{i_{2k-1}}=\epsilon_{i_1i_2\cdots i_{2k-1}}d^{2k-1}x.
\end{align}
Consequently, the $k$th Chern number is expressed as 
\be
c_k=\frac{(-i)^{k-1}}{(2k-1)!2^{k-1}\mathcal{A}(S^{2k-1})}\int_{S^{2k-1}}\text{tr}(-ig^{\dagger}dg)^{2k-1}=(-i)^{k-1}\frac{1}{(2\pi)^k}\frac{(k-1)!}{(2k-1)!} \int_{S^{2k-1}}\text{tr}(-ig^{\dagger}dg)^{2k-1}.
\ee
In low dimensions, we have 
\begin{align}
&c_1=\frac{1}{2\pi}\int_{S^{1}}\text{tr}(-ig^{\dagger}dg),\nn\\
&c_2=-i\frac{1}{24\pi^2} \int_{S^{3}}\text{tr}(-ig^{\dagger}dg)^3,\nn\\
&c_3=-\frac{1}{480\pi^3} \int_{S^{5}}\text{tr}(-ig^{\dagger}dg)^5,\nn\\
&c_4=i\frac{1}{13440\pi^4} \int_{S^{7}}\text{tr}(-ig^{\dagger}dg)^7.
\end{align}
From the general integral expression of $c_k$:   
\be
c_k=\int_{S^{2k-1}}\rho_{2k-1},  
\ee
we define 
\be
\rho_{2k-1}=(-i)^{k-1}\frac{1}{(2\pi)^k}\frac{(k-1)!}{(2k-1)!} \text{tr}(-ig^{\dagger}dg)^{2k-1}, 
\label{defofrho2k-1}
\ee
which satisfies 
\be
d\rho_{2k-1}=0, 
\ee
since $d[\text{tr}(-ig^{\dagger}dg)^{2k-1}]=-\text{tr}(-ig^{\dagger}dg)^{2k}=0.$ 
Due to the Poincar$\acute{\text{e}}$ lemma, $\rho_{2k-1}$ is locally expressed as  
\begin{equation}
\rho_{2k-1}=d\Lambda_{2k-2}. 
\end{equation}
 $\Lambda_{2k-2}$ corresponds to the $U(1)$ transition function of the $(2k-1)$ form  gauge field [see (\ref{u1transc2k-1})]. The associated $U(1)$ topological charge $q_k$ is given by 
\be
q_k\equiv\int_{S^{2k-1}}d\Lambda_{2k-2} = \int_{S^{2k-1}}\rho_{2k-1}=c_k, 
\ee
which is exactly equal to the $k$th Chern number and consistent with (\ref{Ckckequiv}). 

We can also show that the pure gauge Chern-Simons action reproduces  $\rho_{2k-1}$ 
on the equator $S^{2k-1}$. 
The $SO(2k)$ non-abelian gauge fields on north and the south hemispheres are related as  
\begin{equation}
A'=g^{\dagger}Ag-ig^{\dagger}dg, 
\end{equation}
where $g$ is given by  
\begin{equation}
g=\frac{1}{\sqrt{1-{x_{2k+1}}^2}}(x_{2k}+i\gamma_{i}x_{i}). 
\end{equation}
Here,  we used 
\begin{align}
&A=i\frac{1}{2}(1-x_{2k+1})dg g^{\dagger},\nn\\
&A'=-i\frac{1}{2}(1+x_{2k+1})g^{\dagger}dg.
\end{align}
On the equator of $S^{2k}$,  the transition function is reduced to (\ref{mapisoptrans}):  
\begin{equation}
g~\overset{x_{2k+1}=0}\longrightarrow~ x_{2k}+ix_i\gamma_i. 
\end{equation}
In the pure gauge 
\begin{equation}
A=-ig^{\dagger}dg, 
\end{equation}
the Chern-Simons action (\ref{generalexcernsimonsterm}) is reduced to   
\begin{equation}
L_{\text{CS}}^{2k-1}
=(-i)^{k-1}\frac{k!(k-1)!}{(2k-1)!}\text{tr}(-ig^{\dagger}dg)^{2k-1},  
\label{puregaugecsterm}
\end{equation}
where we used 
\begin{equation}
s(k)=k\int_0^1 dt (t-t^2)^{k-1}=\frac{k!(k-1)!}{(2k-1)!}.   
\end{equation}
Thus on the equator $S^{2k-1}$,  the pure Chern-Simons action coincides with the $U(1)$ tensor transition function 
up to a proportional factor:   
\be
L_{\text{CS}}^{2k-1}=i^{2k-1}(2\pi)^k k! ~\rho_{2k-1}. 
\ee

\section{Linking number between membranes}\label{append:linkingmem}

The description here is mainly based on Refs.\cite{TzeNam1989, WuZee1988, NepomechieZee1984}. 
The  tensor Chern-Simons action is given by 
\begin{align}
S&=
-\frac{2}{(2k-1)!}\int d^{4k-1} x ~J_{\mu_1\mu_2\cdots \mu_{2k-1}}C^{\mu_1\mu_2\cdots\mu_{2k-1}}\nn\\
&+\frac{1}{\theta}\frac{1}{(2k-1)!(2k)!}  \int d^{4k-1}x ~\epsilon^{\mu_1\mu_2\cdots\mu_{4k-1}}C_{\mu_1\mu_2\cdots\mu_{2k-1}}G_{\mu_{2k}\mu_{2k+1}\cdots \mu_{4k-1}}.  
\label{actionformembranes} 
\end{align}
In accordance with (\ref{tensorchernsimonslagcoup}), $\theta$ should be taken as     
\be
\theta=2\pi m^k,  
\ee
however in the following we render $\theta$ an arbitrary parameter.  
We derive a higher dimensional Hopf Lagrangian by integrating out the Chern-Simons gauge field.  
The equation for the Chern-Simons field is derived as  
\be
J_{\mu_1\mu_2\cdots\mu_{2k-1}}=\frac{1}{\theta (2k)!}\epsilon_{\mu_1\mu_2 \cdots \mu_{4k-1}}G^{\mu_{2k}\mu_{2k+1}\cdots \mu_{4k-1}}, 
\ee
or 
\be
G^{\mu_1\mu_2\cdots \mu_{2k}}=-\theta\frac{1}{(2k-1)!}\epsilon^{\mu_1\mu_2\cdots \mu_{4k-1}}J_{\mu_{2k+1} \mu_{2k+2}\cdots \mu_{4k-1}}. 
\ee
Since the tensor Chern-Simons field strength is given by (\ref{gfromctensor}), 
it is obvious that the current satisfies a generalized current conservation law: 
\be
\partial^{\mu_i}J_{\mu_1\cdots\mu_i\cdots \mu_{2k-1}}=0 ~~(i=1,2,\cdots,2k-1).
\ee
In a Coulomb like gauge 
$\partial^{\mu}C_{\mu_1\cdots \mu\cdots \mu_{2k-1}}=0$, the Chern-Simons field is expressed as 
\be
C^{\mu_1\mu_2\cdots\mu_{2k-1}}=-\theta\frac{1}{(2k-1)!}\epsilon^{\mu_1\mu_2\cdots \mu_{4k-1}}\partial_{\mu_{2k}}\frac{1}{\partial^2}J_{\mu_{2k+1}\cdots \mu_{4k-1}}, 
\label{afromjmembrane}
\ee
where we used the formula 
\be
\frac{1}{\epsilon\partial}=-\frac{1}{(2k-1)!}\epsilon \partial \frac{1}{\partial^2}.
\ee
By substituting (\ref{afromjmembrane}) to  (\ref{actionformembranes}), we have 
\be
S_{Hopf}=\theta 
\frac{1}{((2k-1)!)^2}\int d^{4k-1}x ~\epsilon^{\mu_1\mu_2\cdots \mu_{4k-1}} J_{\mu_1\mu_2\cdots \mu_{2k-1}}\partial_{\mu_{2k}}\frac{1}{\partial^2} J_{\mu_{2k+1}\cdots \mu_{4k-1}}. 
\label{membraneHopfterm}
\ee 
In the thin membrane limit\footnote{The thin membrane current is given by 
\be
J^{\mu_1\mu_2\cdots\mu_{2k-1}}(x)=\int d^{2k-1}\sigma \frac{\partial(y^{\mu_1},y^{\mu_2},\cdots,y^{\mu_{2k-1}})}{\partial(\sigma^0,\sigma^1,\cdots,\sigma^{2k-2})}\delta^{(4k-1)}(x-y(\sigma)), 
\label{thincurrent}
\ee
where 
\be
\frac{\partial (y^{\mu_1}, y^{\mu_2}, \cdots, y^{\mu_{p+1}})}{\partial (\sigma_0,\sigma_1,\cdots,\sigma_p)}\equiv\epsilon_{\alpha_1\alpha_2\cdots \alpha_{p+1}} \frac{\partial y^{\mu_1}}{\partial \sigma_{\alpha_1}} \frac{\partial y^{\mu_2}}{\partial \sigma_{\alpha_2}}  \cdots    \frac{\partial y^{\mu_{p+1}}}{\partial \sigma_{\alpha_{p+1}}}
\ee
denotes the Jacobian.  },  
(\ref{membraneHopfterm}) yields the linking number of two $(2k-2)$ branes: 
\be
S_{Hopf}~\Rightarrow ~\theta L(V_1,V_2). 
\label{hopfmembrane}
\ee
Here $L(V_1,V_2)$ denotes the higher dimensional generalization of the linking number:  
\be
L(V_1,V_2)=\frac{1}{((2k-1)!)^2 \mathcal{A}(S^{4k-2})} \oint_{V_1} dx^{\mu_1\mu_2\cdots\mu_{2k-1}}\oint_{V_2} d{x'}^{\mu_{2k+1}\mu_{2k+2}\cdots \mu_{4k-1}} \epsilon_{\mu_1\mu_2\cdots \mu_{4k-1}}\frac{x_{\mu_{2k}}(\sigma)-x'_{\mu_{2k}}(\sigma')}{|x(\sigma)-x'(\sigma')|^{4k-1}}, 
  \label{linkingnumbermemb}
\ee
with  
\begin{align}
&dx^{\mu_1 \mu_2 \cdots \mu_{2k-1}}\equiv d\sigma^{0}d\sigma^1\cdots d\sigma^{2k-2} \frac{\partial( x^{\mu_1},x^{\mu_2},\cdots, x^{\mu_{2k-1}})}{\partial(\sigma^0,\sigma^1,\cdots,\sigma^{2k-2})},\nn\\
&dy^{\mu_1 \mu_2 \cdots \mu_{2k-1}}\equiv d{\sigma'}^{0}d{\sigma'}^1\cdots d{\sigma'}^{2k-2} \frac{\partial( y^{\mu_1},y^{\mu_2},\cdots, y^{\mu_{2k-1}})}{\partial({\sigma'}^0,{\sigma'}^1,\cdots,{\sigma'}^{2k-2})}.
\end{align}
With use of the normalized relative coordinates 
\be
z_{\mu}(\sigma,\sigma')\equiv \frac{x_{\mu}(\sigma)-x_{\mu}'(\sigma')}{|x(\sigma)-x'(\sigma')|}, 
\ee
the the linking number (\ref{linkingnumbermemb}) is concisely expressed as 
\be
L(V_1,V_2)=\frac{1}{(4k-2)!\mathcal{A}(S^{4k-2})} \int d{z}^{\mu_1}d{z}^{\mu_2}\cdots d{z}^{\mu_{4k-2}} \epsilon_{\mu_1\mu_2\cdots \mu_{4k-1}} {z}_{\mu_{4k-1}}, 
\label{zthinlinkmem}
\ee
where 
\be
d{z}^{\mu_1}d{z}^{\mu_2}\cdots d{z}^{\mu_{4k-2}}\equiv d\sigma_0 d\sigma_1\cdots d\sigma_{2k-2} d\sigma'_0 d\sigma'_1\cdots d\sigma'_{2k-2} \frac{\partial({z}^{\mu_1},{z}^{\mu_2},\cdots, {z}^{\mu_{4k-2}}) }{\partial(\sigma_0,\cdots,\sigma_{2k-2},\sigma'_0,\cdots,\sigma'_{2k-2})}. 
\ee
Here, we used the formula of the determinant (\ref{detformula2k})
\footnote{
In the map from a set of $2k$ coordinates, $\sigma_1,\sigma_2,\cdots,\sigma_{2k}$, to another set of $2k$ coordinates, $z_{\mu}=z_{\mu}(\sigma_1,\sigma_2,\cdots, \sigma_{2k})$  $(\mu={1,2,\cdots,2k})$, the volume density is given by   
\be
D\equiv \frac{\partial(z_1,z_2,\cdots, z_{2k})}{\partial(\sigma_1,\sigma_2,\cdots,\sigma_{2k})}\equiv \epsilon_{\mu_1\mu_2\cdots\mu_{2k}} \frac{\partial z_{\mu_1}}{\partial \sigma_1} \frac{\partial z_{\mu_2}}{\partial \sigma_2}\cdots  \frac{\partial z_{\mu_{2k}}}{\partial \sigma_{2k}}, 
\ee
which can be rewritten as 
\be
D=\frac{1}{(2k)!}\epsilon_{\mu_1\mu_2\cdots \mu_{2k}} \frac{\partial(z_{\mu_1},z_{\mu_2},\cdots, z_{\mu_{2k}})}{\partial(\sigma_1,\sigma_2,\cdots,\sigma_{2k})} =\frac{1}{(k!)^2}\epsilon_{\mu_1\mu_2\cdots \mu_{2k}} \frac{\partial(z_{\mu_1},z_{\mu_2},\cdots, z_{\mu_{k}})}{\partial(\sigma_1,\sigma_2,\cdots,\sigma_{k})} \frac{\partial(z_{\mu_{k+1}},z_{\mu_{k+2}},\cdots, z_{\mu_{2k}})}{\partial(\sigma_{k+1},\sigma_{k+2},\cdots,\sigma_{2k})}. 
\label{detformula2k}
\ee
}. 
It should be noticed that the integral in (\ref{zthinlinkmem})
\be
\frac{1}{(4k-2)!}
 \int d{z}^{\mu_1}d{z}^{\mu_2}\cdots d{z}^{\mu_{4k-2}} \epsilon_{\mu_1\mu_2\cdots \mu_{4k-1}} {z}_{\mu_{4k-1}}
\ee
 represents the area of $S^{4k-2}$ with coordinates ${z}_{\mu}$ ($\sum_{\mu=1}^{4k-1}z_{\mu}z_{\mu}=1)$.  
 Thus, the linking number (\ref{zthinlinkmem}) can  alternatively be understood as the winding number from the world-volumes of two $(2k-2)$-branes to $S^{4k-2}$: 
\be
(S^{2k-2}\times S^1)\times (S^{2k-2}\times S^1)~~\rightarrow S^{4k-2}.  
\label{higehrlinkwindsphere}
\ee
For $k=1$, (\ref{linkingnumbermemb}) is reduced to  the original Gauss linking \cite{Gauss1833,Maxwell1873,AshtekarCorichi1997}: 
\be
L(C_1,C_2)=\frac{1}{4\pi} \oint_{C_1} dx^{\mu}\oint_{C_2} d{x'}^{\rho} \epsilon_{\mu\nu\rho}\frac{x_{\nu}(\sigma)-x'_{\nu}(\sigma')}{|x(\sigma)-x'(\sigma')|^{3}},  
\ee
and similarly (\ref{higehrlinkwindsphere}) becomes to   
\be
T^2\equiv S^1\times S^1 ~~\rightarrow~~S^2.  
\ee




\end{document}